\newif\ifjournal  % JOURNAL SUBMISSION?
\newif\ifblind    % BLIND
\newif\ifamd      % include AMD results

\amdtrue

\ifjournal
  \ifblind
    \documentclass[sigconf,review,anonymous]{acmart}
  \else
    \documentclass[sigconf]{acmart}
  \fi

  \copyrightyear{2022}
  \acmYear{2022}
  \setcopyright{licensedusgovmixed}\acmConference[ICPP '22]{51st International Conference on Parallel Processing}{August 29-September 1, 2022}{Bordeaux, France}
  \acmBooktitle{51st International Conference on Parallel Processing (ICPP '22), August 29-September 1, 2022, Bordeaux, France}
  \acmPrice{15.00}
  \acmDOI{10.1145/3545008.3546185}
  \acmISBN{978-1-4503-9733-9/22/08}

  \definecolor{RedOrange}{rgb}{0.8, 0.33, 0.0}
  \definecolor{Cerulean}{rgb}{0.61, 0.77, 0.89}
  \definecolor{OliveGreen}{rgb}{0.33, 0.42, 0.18}
  \definecolor{Plum}{rgb}{0.56, 0.27, 0.52}
\else
  \documentclass[twoside,leqno,twocolumn]{article}

  \usepackage{orcidlink}

  \usepackage{breakcites}
  \usepackage[square]{natbib}
  \renewcommand\cite{\citep}
  \setcitestyle{aysep={}}

  \usepackage[T1]{fontenc}
  \usepackage[tt=false, type1=true]{libertine}
  \usepackage[varqu]{zi4}
  \usepackage[libertine]{newtxmath}
\fi

\usepackage{tikz} % FIXME TAPS
\usetikzlibrary{arrows,positioning,shapes}

\usepackage{listings}
\usepackage{algorithm}
\usepackage{amsmath}
\usepackage[noend]{algpseudocode} % FIXME TAPS
\usepackage{amsfonts}
\usepackage[noabbrev]{cleveref} % FIXME TAPS
\usepackage{xspace}
\usepackage{subfig}

\DeclareMathOperator*{\argmin}{arg\,min}

\makeatletter
\def\blfootnote{\xdef\@thefnmark{}\@footnotetext}
\makeatother

\newcommand{\kdtree}{\textit{k}-d tree\xspace}
\newcommand{\boruvka}{Bor\r{u}vka\xspace}
\newcommand{\hdbscan}{\textsc{Hdbscan*}\xspace}
\newcommand{\emst}{\textsc{Emst}\xspace}
\newcommand{\mst}{\textsc{Mst}\xspace}
\newcommand{\wspd}{\textsc{Wspd}\xspace}
\newcommand{\bcp}{\textsc{Bcp}\xspace}

\newcommand{\bigo}[1]{\mathcal{O}\!\left( #1 \right)}

\newcommand{\nvidiagpu}{Nvidia A100\xspace}
\newcommand{\amdgpu}{AMD MI250X\xspace}
\newcommand{\amdcpu}{AMD EPYC 7763\xspace}
\newcommand{\dataset}[1]{\emph{#1}\xspace}
\newcommand{\wangemst}{\textsc{MemoGFK}\xspace}
\newcommand{\arborx}{\textsc{ArborX}\xspace}
\newcommand{\mlpack}{\textsc{MLPACK}\xspace}

\newcommand{\MilFeatPerSec}[1]{#1 \emph{MFeatures/sec}\xspace}

\begin{document}

\title{A single-tree algorithm to compute the Euclidean minimum spanning tree on GPUs}

\ifjournal
  \unless\ifblind
    \author{Andrey Prokopenko}
    \email{prokopenkoav@ornl.gov}
    \orcid{0000-0003-3616-5504}
    \affiliation{%
      \institution{Oak Ridge National Laboratory}
      \streetaddress{1 Bethel Valley Rd}
      \city{Oak Ridge}
      \state{Tennessee}
      \country{USA}
      \postcode{37830}
    }
    \author{Piyush Sao}
    \orcid{0000-0002-9432-5855}
    \affiliation{%
      \institution{Oak Ridge National Laboratory}
      \streetaddress{1 Bethel Valley Rd}
      \city{Oak Ridge}
      \state{Tennessee}
      \country{USA}
      \postcode{37830}
    }
    \author{Damien Lebrun-Grandi\'e}
    \orcid{0000-0003-1952-7219}
    \affiliation{%
      \institution{Oak Ridge National Laboratory}
      \streetaddress{1 Bethel Valley Rd}
      \city{Oak Ridge}
      \state{Tennessee}
      \country{USA}
      \postcode{37830}
    }

    \renewcommand{\shortauthors}{Prokopenko et al.}
  \fi
\else
  \author{
    A.~Prokopenko\thanks{Oak Ridge National Laboratory}\enskip\orcidlink{0000-0003-3616-5504},
    P.~Sao\footnotemark[1]\enskip\orcidlink{0000-0002-9432-5855},
    D.~Lebrun-Grandi\'e\footnotemark[1]\enskip\orcidlink{0000-0003-1952-7219}
  }
  \date{}
\fi

\ifjournal
  \begin{abstract}
    Computing the Euclidean minimum spanning tree (\emst) is a computationally
demanding step of many algorithms. While work-efficient serial and
multithreaded algorithms for computing \emst are known, designing an efficient
GPU algorithm is challenging due to a complex branching structure, data
dependencies, and load imbalances. In this paper, we propose a single-tree
\boruvka-based algorithm for computing \emst on GPUs. We use an efficient
nearest neighbor algorithm and reduce the number of the required distance
calculations by avoiding traversing subtrees with leaf nodes in the same
component. The developed algorithms are implemented in a performance portable
way using ArborX, an open-source geometric search library based on the Kokkos
framework. We evaluate the proposed algorithm on various 2D and 3D datasets,
show and compare it with the current state-of-the-art open-source CPU
implementations. We demonstrate 4-24$\times$ speedup over the fastest multi-threaded
implementation. We prove the portability of our implementation by providing
results on a variety of hardware: \amdcpu, \nvidiagpu and \amdgpu. We show
scalability of the implementation, computing \emst for 37~million 3D
cosmological dataset in under a 0.5 second on a single A100 Nvidia GPU.

  \end{abstract}
  % FIXME: FIgure this out

\fi

\maketitle

\unless\ifjournal
  Computing the Euclidean minimum spanning tree (\emst) is a computationally
demanding step of many algorithms. While work-efficient serial and
multithreaded algorithms for computing \emst are known, designing an efficient
GPU algorithm is challenging due to a complex branching structure, data
dependencies, and load imbalances. In this paper, we propose a single-tree
\boruvka-based algorithm for computing \emst on GPUs. We use an efficient
nearest neighbor algorithm and reduce the number of the required distance
calculations by avoiding traversing subtrees with leaf nodes in the same
component. The developed algorithms are implemented in a performance portable
way using ArborX, an open-source geometric search library based on the Kokkos
framework. We evaluate the proposed algorithm on various 2D and 3D datasets,
show and compare it with the current state-of-the-art open-source CPU
implementations. We demonstrate 4-24$\times$ speedup over the fastest multi-threaded
implementation. We prove the portability of our implementation by providing
results on a variety of hardware: \amdcpu, \nvidiagpu and \amdgpu. We show
scalability of the implementation, computing \emst for 37~million 3D
cosmological dataset in under a 0.5 second on a single A100 Nvidia GPU.

\fi

\unless\ifblind
  \blfootnote {%
  % ORNL disclaimer
  This manuscript has been authored by UT-Battelle, LLC, under contract
  DE-AC05-00OR22725 with the U.S. Department of Energy. The United States
  Government retains and the publisher, by accepting the article for publication,
  acknowledges that the United States Government retains a nonexclusive, paid-up,
  irrevocable, world-wide license to publish or reproduce the published form of
  this manuscript, or allow others to do so, for United States Government
  purposes.
  }
\fi

% FIXME remove this before submission
% \nocite{*}

\section{Introduction}\label{s:introduction}
\begin{figure}
  \includegraphics[width=\columnwidth]{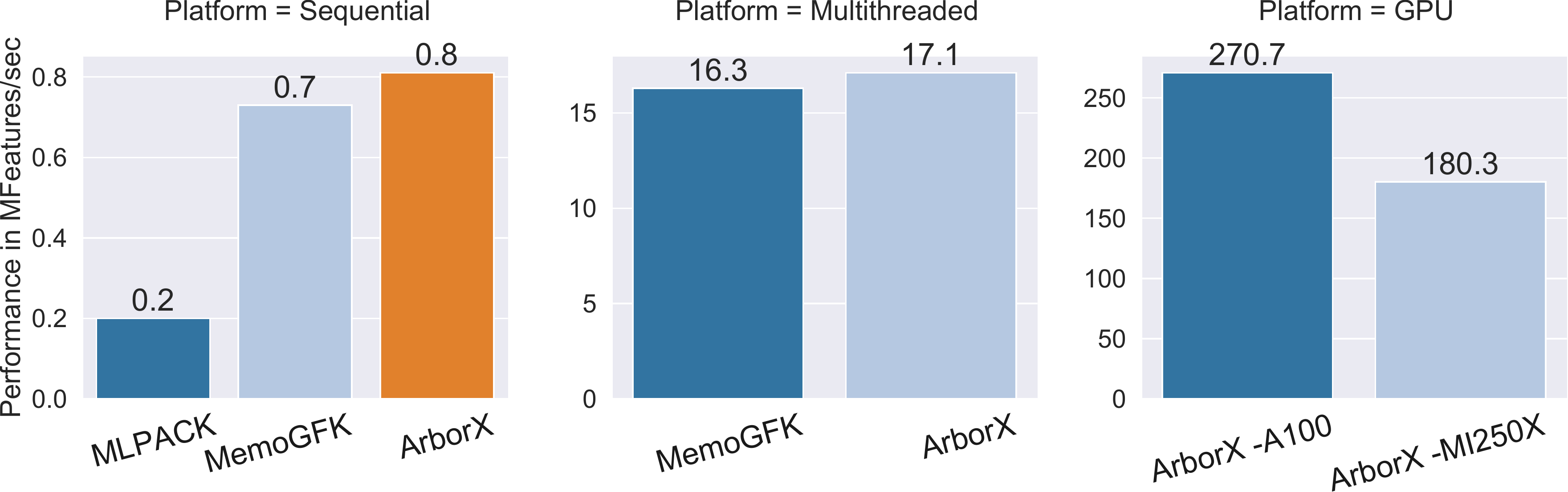}
  \caption{\label{fig:perfSummary} Performance (in\MilFeatPerSec{}) for the
  dual-tree~\cite{march2010dualtreeMST} (MLPACK), WSPD~\cite{wang2021fast}
  (\wangemst), and single-tree (\textbf{this work using \arborx}) approaches on
  \amdcpu CPU (sequential and multi-threaded), and \nvidiagpu and \amdgpu
  (single GCD) GPU architectures for a 3D cosmological dataset (\dataset{Hacc37M}).}
\end{figure}
% The \emph{minimum spanning tree} (MST) problem is a common graph optimization
% problem. It is stated as follows: given a connected undirected graph with
% real-valued weights assigned to each edge, find a spanning tree of the graph
% with a minimum total edge cost. Almost all algorithms to compute MST are a
% variation of the following generic algorithm. Starting from a forest of
% one-vertex trees, the generic algorithm connects the trees by choosing a
% suitable edge, maintaining an acyclic subgraph of the input graph. At any step,
% this subraph is a part of the final MST. At the end of the procedure, a single
% spanning tree is constructed.

Given a set of $n$ points in a $d$-dimensional space, the \emph{Euclidean
minimum spanning tree} (\emst) problem determines the minimum spanning tree
(\mst) of the distance graph of the set, i.e., a graph where each pair of
vertices are connected by an edge of weight equal to the Euclidean distance
between them. Computing \emst is an important task in a variety of
applications, including data clustering~\cite{campello2015hdbscan}, Euclidean
traveling salesman problem~\cite{held1970traveling},
cosmology~\cite{naidoo2020beyond}, wireless network
connectivity~\cite{li2001constructing}, computational fluid
dynamics~\cite{subramaniam1998mixing}, and many others.

% All algorithms to compute MST are based on a greedy approach. At any instant
% during the computation, the \mst~algorithm maintains a forest, e.g. collection
% of subtrees and disconnected vertices. In each step, the algorithm connects some
% of the subtrees while maintaining the forest structure. The algorithm terminates
% when only a tree is left for each connected component in the graph.

\begin{figure*}[h]
  \subfloat[Initial state]{\includegraphics[width=0.17\textwidth]{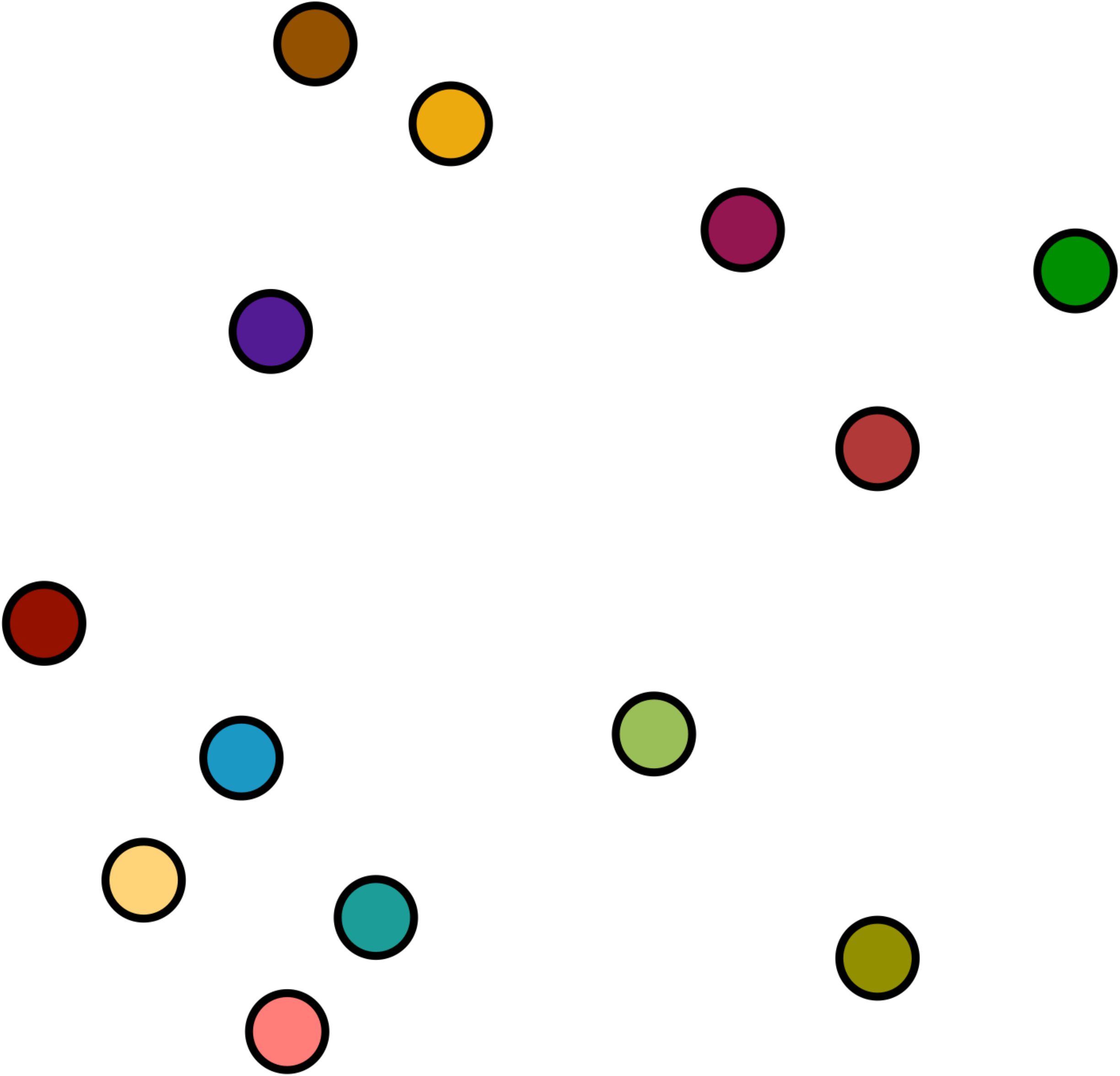}}
  \hfill
  \subfloat[Components at some iteration\label{f:boruvka_step2}]{\includegraphics[width=0.17\textwidth]{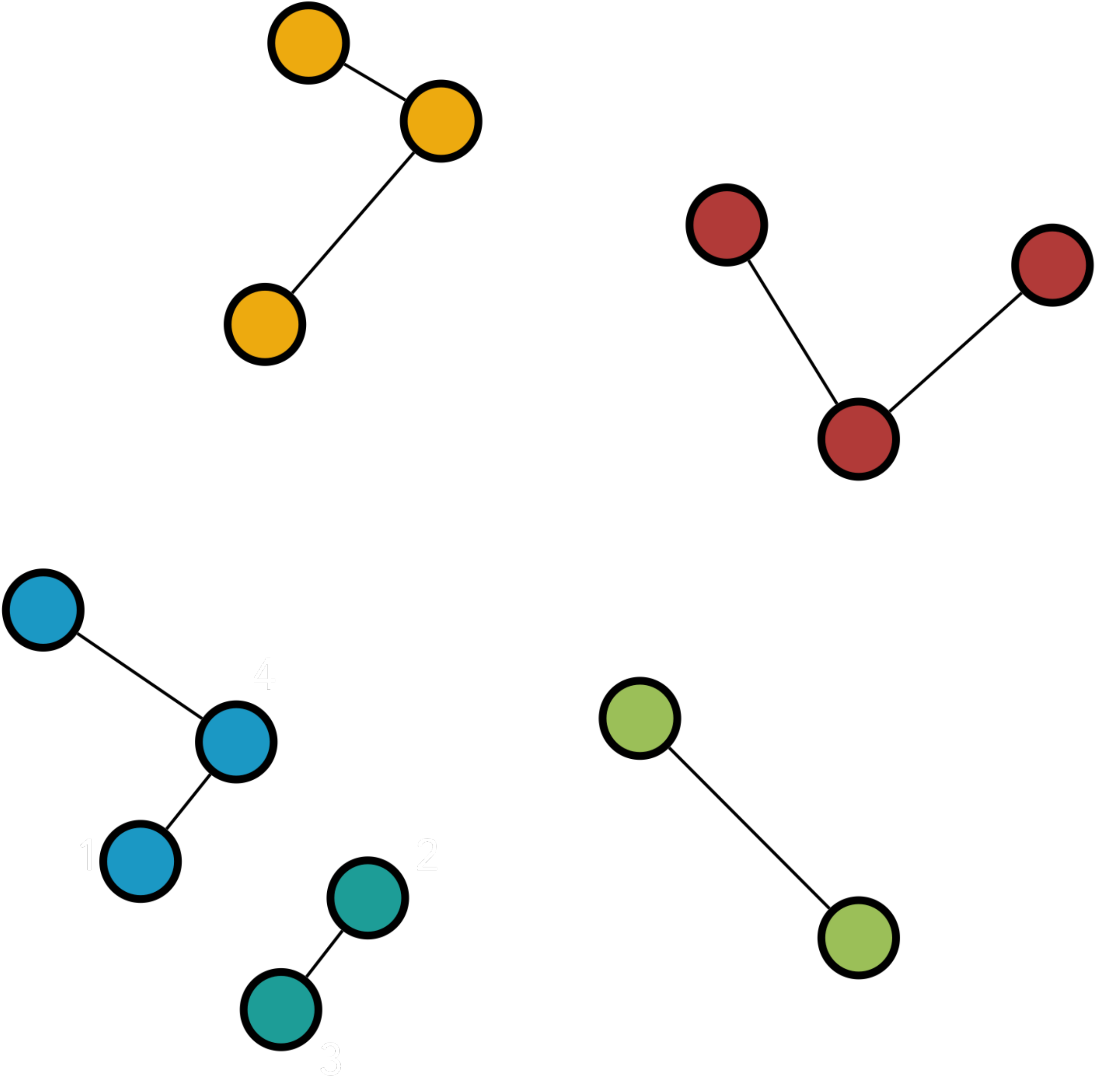}}
  \hfill
  \subfloat[Candidate edges\label{f:boruvka_step3}]{\includegraphics[width=0.17\textwidth]{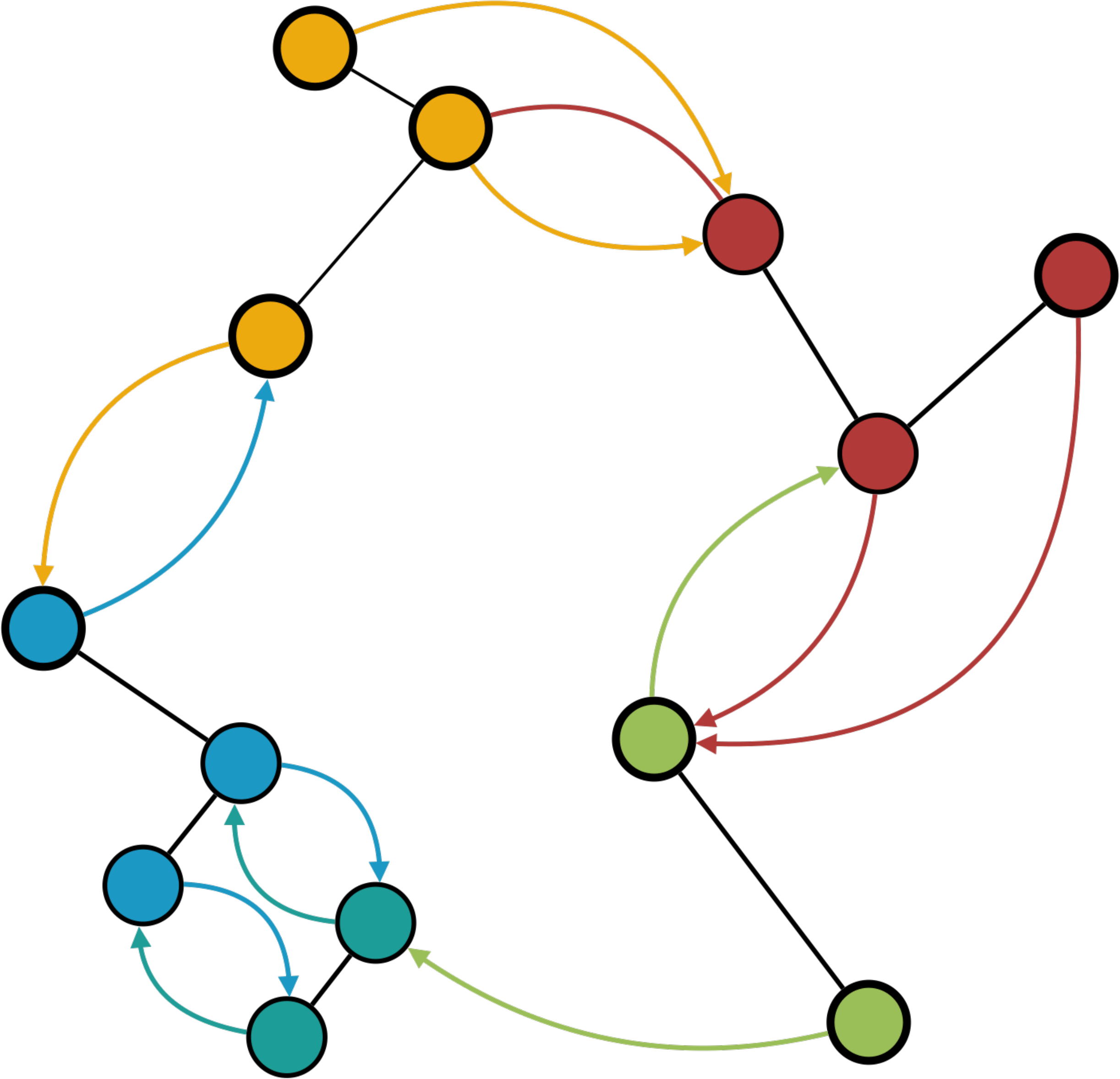}}
  \hfill
  \subfloat[Closest components\label{f:boruvka_step4}]{\includegraphics[width=0.17\textwidth]{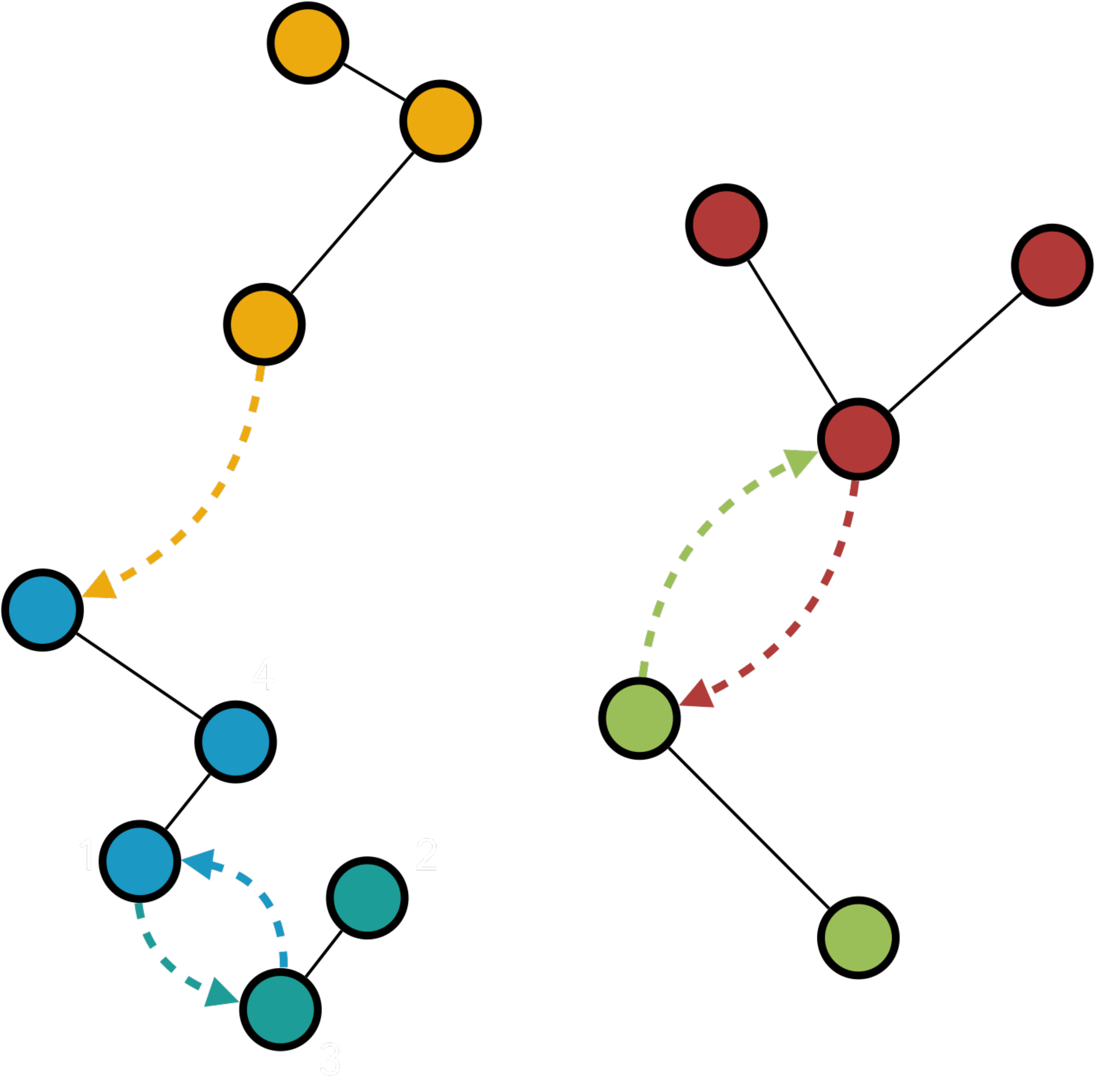}}
  \hfill
  \subfloat[New components after the merge]{\includegraphics[width=0.17\textwidth]{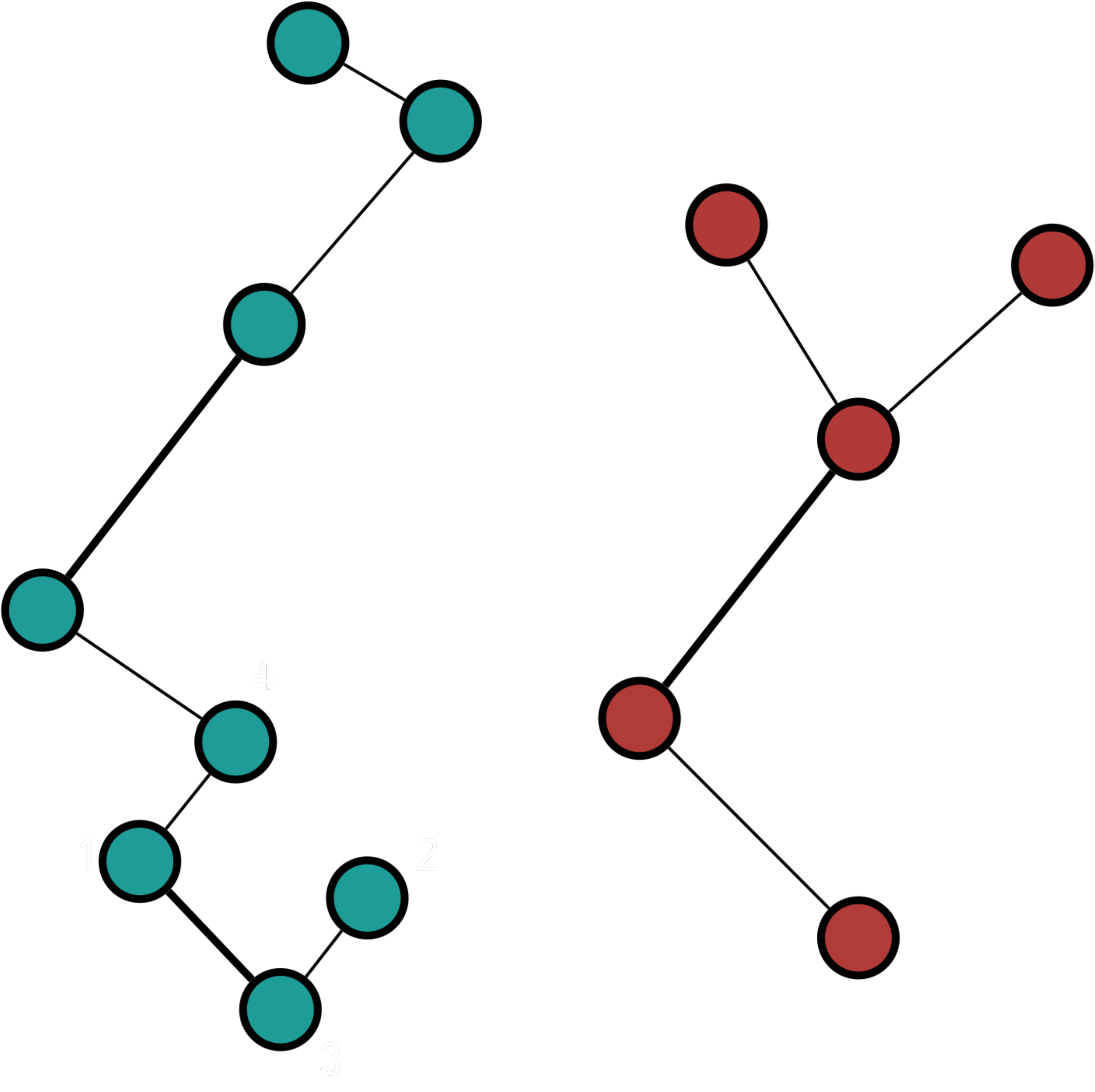}}
  \caption{\label{fig:boruvka}\boruvka's algorithm.
  (a) Initial state (each component having a single vertex).
  (b) The state after a few \boruvka iterations.
  (c) Closest neighbors from a different component for each vertex.
  (d) The shortest outgoing edge for each component.
  (e) The new components after the merge (the initial state of the next \boruvka iteration).
  }
\end{figure*}

Most algorithms for computing an \mst~on a general graph are variants of the
three classical algorithms: 1926 \boruvka's algorithm~\cite{boruvka1926}, 1956
Kruskal's algorithm~\cite{kruskal1956} and 1957 Prim's
algorithm~\cite{prim1957}. These algorithms share the same general idea,
constructing an \mst iteratively. At any instant during the computation, an
algorithm maintains a set of non-overlapping sets of vertices called
components. Initially, all components consist of a single vertex. On each step
of an algorithm, some components are merged using a subset of the graph edges.
The algorithm terminates when there remain no edges connecting separate
components.
\iffalse
The complexity of these classical algorithms is linear in the
number of edges in the graphs. \fix{AP}{Is that really true? I thought
$O(mlogn)$. I'd rather remove this sentence.}
\fi

The fundamental difference of the \emst problem and the \mst one lies in the
graph structure. \mst operates on a sparse graph, where the number of edges is
a small fraction of all possible edges with the same vertices. On the other
hand, \emst uses the distance graph, which is \emph{complete}, with each pair
of vertices connected by an edge, for a total of $n(n-1)/2$ edges. It is
prohibitively expensive to both construct and store a complete graph for large
problems, as well as run classical \mst algorithms on such a graph. Thus, the
distance graph is typically used implicitly.

To solve the \emst problem, Bentley and Friedman~\cite{bentley1978} proposed
combining classical \mst algorithms with a nearest-neighbor search algorithm
using a spatial indexing structure (e.g., a \kdtree). However, a
straightforward implementation of this approach performs poorly. The main
bottleneck of the algorithm is in the excessive number of distance calculations
in the later iterations of an \mst algorithm. Since, the key challenge in
designing an efficient \emst algorithm has been in careful pruning of distance
calculations during the runtime.

Two popular approaches have emerged: one based on the \emph{well separated pair
decomposition} (\wspd)~\cite{callahan1995wspd,narasimhan2000emst,wang2021fast},
and the other using the \emph{dual-tree} framework~\cite{march2010dualtreeMST}.
Under some mild assumptions on the distribution of the data points, the
dual-tree method provides the sharpest worst-case bounds for any dimensional
space. The two approaches have been shown to perform well on CPUs, including
multi-threaded parallelization.

However, the existing approaches are limited in scalability and performance for
an efficient GPU computation.
% The tree building phase is unscalable, even on a
% multithreaded architecture.
The dual-tree algorithm, in general, is outperformed by the best-known
\wspd-based approach. And while \wspd is asymptotically
$\bigo{n}$\cite{agarwal1991bcp}, the hidden constants of the algorithm are very
high. In fact, we observed that the \wspd computation dominates the overall
time in the \emst computation  and is comparable with a single-tree
implementation in the sequential case. Thus, an \emst algorithm or
implementation that is sequentially efficient, is scalable with respect to
problem size, and is amenable to GPU parallelism remains an open problem.

% Realizing high-performance GPU implementation of the
% known work optimal sequential and multithreaded parallel algorithm for computing
% \emst is challenging as they exhibit complex branching structure and data
% dependencies, and suffer from load imbalances.

% The focus of this paper is to describe an efficient algorithm to compute \emst
% on a GPU.  In this paper, we present a
% high-performance \emph{single}-tree  Barouvka algorithm for computing \emst on
% GPUs.

In this paper, we propose an efficient algorithm for \emst suitable for both
CPUs and GPUs. Our algorithm is based on \boruvka's algorithm and uses a
\emph{single} tree to perform the nearest-neighbor queries. We use a bounding
volume hierarchy (BVH) as our tree structure as it is very efficient for
unstructured low-dimensional data on GPU (we note, however, that the described
algorithms are general and are applicable to other tree structures such as
\kdtree). To reduce the number of the distance calculations when finding the
closest outgoing edge for a given component (the most expensive part of each
\boruvka iteration), we keep track of the component membership of the children of
the internal tree nodes. This allows nearest neighbor queries to bypass subtrees
where all leaf nodes lie in the same component.
% During the tree traversal, threads corresponding to points within the same
% component may share a common upper bound for the closest neighbor distance,
% allowing earlier traversal termination for many threads.

Our motivation for this work comes from an astronomy application which requires
high performance to analyze the data from a cosmological simulation. We show
the results from one such dataset in~\Cref{fig:perfSummary}. Additionally, we
show that our GPU implementation achieves 270\MilFeatPerSec{} (million
features, the product of the number of points and dimensions, processed per
second) on nVidia A100, which is 17$\times$ faster than best known multithreaded
implementation.

In our implementation, we used ArborX~\cite{arborx2020}, a performance portable
geometric search library using Kokkos framework~\cite{kokkos2022}. This allows
us to study the algorithm on both CPU and multiple GPU architectures (e.g.,
\nvidiagpu, \amdgpu). We evaluate the proposed algorithm on various 2D and 3D
datasets, and compare it with the current state-of-the-art open-source CPU
implementations.

Our key contributions are:
\begin{itemize}
\item
  We provide the first performance portable algorithm and implementation for
  the  \emst problem. Our algorithm is efficient in the sequential case and
  outperforms the best publicly known sequential algorithm in a best-case by
  50\% (see \Cref{sec:serial-perf}).
\item
  Compared to the best available multithreaded algorithm, our GPU
  implementation is up to 24$\times$ faster, and our multi-threaded
  implementation is within 0.5-2$\times$. This results in a best-case
  performance of \MilFeatPerSec{270}. In contrast, the
  best sequential algorithm achieves \MilFeatPerSec{1.2} on the latest
  \amdcpu and \MilFeatPerSec{16} by an efficient-multithreaded
  implementation(See~\cref{sec:parallel-perf}).

\item
 We show that our proposed algorithm gracefully handles certain non-Euclidean
 distances. Specifically, we show that our algorithm is efficient when used
 with \emph{mutual reachability distance}, a variant of Euclidean distance
 used in a popular clustering algorithm \hdbscan~\cite{mcinnes2017hdbscan}.

\item
  We provide a comprehensive set of experiments on three architectures to
  establish or provide empirical evidence for several properties of our
  algorithm and implementation including performance portability, asymptotic
  linear cost growth with problem size and lower threshold problem size to
  achieve performance saturation relative to CPU.
\end{itemize}
% We show scalability of the implementation, computing \emst for
% 37M million 3D points in under a 0.5 second on a single \nvidiagpu GPU.

% Additionally, we show how our \emst algorithm can be adapted to certain
% non-Euclidean distance metrics (e.g., \emph{mutual-reachability} distance),
% used in clustering algorithms such as \hdbscan~\cite{mcinnes2017hdbscan}. This
% only requires a minor change in the tree traversal algorithm.

The remainder of the paper is organized as follows. \Cref{s:background}
introduces the \boruvka algorithm and gives an overview of the related work.
\Cref{s:algorithm} describes the proposed algorithm. We compare our
implementation with the state-of-the-art CPU implementations and demonstrate
the algorithm's performance on GPUs in \Cref{s:results}.

% The algorithms vary in their choice of the trees to connect. If only two trees
% can be connected at a given step without affecting the rest of the algorithm,
% this necessarily leads to a serial algorithm. For an algorithm to be suitable
% for parallel computations, multiple trees must be connected at the same time
% without affecting the overall result.

% \boruvka's algorithm~\cite{boruvka1926} (and its variants \improve{AP}{explore
% literature on different variants of parallel \boruvka's algorithms}) exhibit
% desired parallel characteristics. The vertices in each tree can be seen as
% disjoint sets component. Each such component defines a \emph{cut}, a set of all
% edges with one vertex inside the component, and one outside. At each step, the
% algorithm selects the edge with the smallest weight among all edges of a cut of
% each component, and connects the trees that the vertices of that edge belong to.
% The algorithm converges in at most $\log_2(n)$ steps, as on each step each
% component is connected to at least one other.

\section{Background}\label{s:background}
% \begin{figure}
% \centering
% \includegraphics[width=0.95\columnwidth]{figs/inout.pdf}
% \caption{ \label{fig:inout} Input and output of \emst computation}
% \end{figure}

In this Section, we will briefly review the background and the related work
relevant to subsequent discussion. \textbf{Notations:} Let $G = \left\{ V,E,
W\right\}$ be a weighted undirected connected graph. Here, $V$ is the set of
vertices of size $n$, $E$ is the set of edges of size $m$, and $W$ are the
weights of the edges. We will use a \emph{component} to describe a subset of
of $V$ with its meaning clear from the context.

%% MST

\textbf{MST computation.} For a connected graph, the \mst is the tree subgraph
with the least sum of edge weights. \mst is unique if all edge weights are
distinct.
Many algorithms to compute \mst are based on a greedy approach, using the fact that the
minimum weight edge in any edge cut will be in \mst if it is unique (if it is
not unique, any one of the edges with minimum weight can be chosen).
The three most popular algorithms for computing \mst are \boruvka's
algorithm~\cite{boruvka1926}, Prim's algorithm~\cite{prim1957}, and Kruskal's
algorithm~\cite{kruskal1956}.

Prim's algorithm operates on a single component. At the start, the component is
assigned a single vertex. On each step of the algorithm, the component is
expanded by adding a vertex connected by an edge of the minimum weight in the
component's cut. Prim's algorithm has $\bigo{m\log n}$ complexity, and is
inherently sequential.

In the Kruskal's algorithm, each vertex is initially assigned to its own
component, and all edges are sorted by weights. On each step, the edge with the
minimum weight is chosen among all edges that have the vertices in different
components. The components of the vertices of that edge are then merged
together. Kruskal's algorithm has $\bigo{m\log n}$ complexity. It allows for
a limited parallelism which is insufficient for a GPU.

%% Boruvka

\textbf{\boruvka's algorithm.} \boruvka's algorithm was one of the first
published algorithms to compute an \mst. Similarly to the Kruskal's algorithm,
it maintains a set of components, each initially containing a single vertex.
Similarly to the Prim's algorithm, each component is expanded through finding
the minimum weight edge in its cut. Unlike Prim's algorithm, however, the
computation of the minimum weight edges can be done in parallel, and components
are expanded through merging components together, rather than adding a single
vertex.

\begin{algorithm}[t]
  \caption{\boruvka's algorithm\label{a:boruvka}}
\begin{algorithmic}[1]
\small
\Procedure{Boruvka}{$G = \{V, E, W\}$}
  \State $T \gets (V, \emptyset)$ \Comment{initialize graph $T$ with vertices from $V$ and no edges}
  \While{$T$ has more than one connected component}
    \ForAll{components $C$ of $T$}
      \State $S \gets \emptyset$
      \ForAll{vertices $v$ in $C$}
        \State $D \gets \{a \in E \; | \; a = (v, w), w \notin C\}$
        \State $e \gets$ minimum weight edge in $D$
        \State $S \gets S \cup {e}$
      \EndFor
      \State $e \gets$ minimum weight edge in $S$
      \State Add $e$ to the graph $T$
    \EndFor
  \EndWhile
\EndProcedure
\end{algorithmic}
\end{algorithm}

At the start of the \boruvka's algorithm (\Cref{a:boruvka}), each component is initialized with an
individual vertex, $C_i = \{v_i\}$. On each step, the
algorithm determines the edge with the minimum weight in the cut of each active
component. In other words, for a component $C_i$ we find an
edge $e_i = (v, u) \in E$, $v \in C_i$ and $u \in
C_j, j \neq i$, with the minimum weight. We will call such edge $e_i$
the \emph{smallest outgoing edge} for the component $C_i$, and denote $C_i
\rightarrow C_j$. The found edge is added to the list of edges in the \mst, and
the two components $C_i$ and $C_j$ are merged, $C_i \gets C_i \cup
C_j$.

It is not guaranteed that the smallest outgoing edge for $C_i$ with an end in
$C_j$ would be the smallest outgoing edge for $C_j$. Instead, the found edges
result will typically produce a chain of components $C_{i_1} \rightarrow \cdots
\rightarrow C_{i_{s-1}} \leftrightarrow C_{i_s}$. Each such chain terminates in
a pair of components with their smallest outgoing edges pointing to each other.
All components belonging to the same chain can be merged together in the
same \boruvka iteration. In practice, this results in the \boruvka's algorithm
requiring far fewer iterations compared to its theoretical upper bound of
$\lceil \log_2(n) \rceil$.

\Cref{fig:boruvka} demonstrates the steps in a single \boruvka
iteration. At the beginning of the $k$-th iteration, we have five components
(\Cref{f:boruvka_step2}). First, each point finds the closest neighbor
belonging to a different component than its own (\Cref{f:boruvka_step3}). This
forms a set of candidate edges for each component. Then, we choose the shortest
candidate edge for each component (\Cref{f:boruvka_step4}) and add it to a set
of found \mst edges. Finally, the newly found \mst edges connect  previously
disconnected components. To merge the components, we compute the new component
label for each point. We can see that one of the new components was formed by
merging three components.

\boruvka's algorithm is guaranteed to converge (i.e., produce a correct \mst)
only when all edge weights are distinct. Otherwise, the found edges may result
in a cycle. This situation may be avoided by a suitable tie-breaking resolution
when selecting the smallest outgoing edges. One of the ways to achieve that is
by using indices of the vertices for the comparison of the edges. For example,
given two edges $e_1 = (v_1, w_1)$ and $e_2 = (v_2, w_2)$ of the same weight,
one could define $e_1 < e_2$ if $\min(v_1, w_1) < \min(v_2, w_2)$, or
$\min(v_1, w_1) = \min(v_2, w_2)$ and $\max(v_1, w_1) < \max(v_2, w_2)$.

The parallel nature of the \boruvka's algorithm make it well suited for a
GPU implementation.

%% EMST

\textbf{\emst computation.} Given a set of points $X$ in a $d$-dimensional
space, Euclidean minimum spanning tree (\emst) is defined as an \mst of its
\emph{distance} graph. The distance graph $\mathcal{D}$ of $X$ is a complete
graph, with each vertex corresponding to a point in $P$, and each edge $e_{ij}
= (p_i, p_j)$ having the weight $w_{ij} = \|p_i - p_j\|_2$. Explicitly
computing and storing $\mathcal{D}$ is undesirable as it requires $\bigo{n^2d}$
operations and $\bigo{n^2}$ storage, which is prohibitively expensive for large
datasets. For that reason, it is usually used implicitly.

As the complexity of the \mst algorithms is at least linear in the number of edges,
regular \mst algorithms are not suitable for the \emst problem as they would
have quadratic complexity with respect to the number of points $\bigo{n^2}$.

For the two-dimensional case, \mst calculation can be performed on a
\emph{Delaunay} triangulation of the points, which only has $\bigo{n}$ edges.
However, Delaunay triangulation worst-case complexity grows from
$\bigo{n\log n}$ in the two-dimensional case to $\Theta\left ( n^{2} \right ) $
for higher dimensions.

Instead, \emst algorithms combine a general \mst algorithm with a data
structure to accelerate the search for the nearest neighbors. Bentley and
Friedman~\cite{bentley1978} proposed the first such \emst algorithm using a
\kdtree-based nearest neighbor searches together with Prim's algorithm. The
authors estimated $\bigo{n\log n}$ operations for most distributions of points,
albeit not rigorously. A key limitation of this approach is that it will often
perform many redundant distance computations. This stems from the iterative
nature of \mst algorithms, where the nearest-neighbor queries can be run
multiple times for the same points.

Pruning the number of the redundant distance computations for \emst was
explored in many  works.
% So in subsequent nearest-neighbor queries, we may be able to
% avoid a lot of distance computations either by suitably caching previously
% computed distance or by leveraging the structure of \boruvka iterations.
The two popular strategies emerged: the \emph{well-separated pair
decomposition} (\wspd)~\cite{callahan1995wspd}, and the \emph{dual-tree}
algorithms~\cite{march2010dualtreeMST}.

A pair of sets of points $(P, Q)$ is called \emph{well-separated} if the
shortest distance between any point in $P$ to any point in $Q$ is greater than
the diameter of both of the sets. For a given set of points, \wspd is defined
as a sequence of well-separated pairs $\left(P_{i}, Q_{i}\right)$ such that for
any pair of points $p, q \in X$ there exists a well-separated pair $(P_k, Q_k)$
with $p \in P_k$ and $q \in Q_k$.
With \wspd, \emst computation can be reformulated as a computation of the
\emph{bichromatic closest pair} (\bcp)~\cite{agarwal1991bcp} between the
well-separated pairs, and performing an \mst computation using the found \bcp edges.
The first algorithm based on this approach was proposed in Agrawal et.
al.~\cite{agarwal1991bcp}. Narasimhan~\cite{narasimhan2000emst} proposed
\emph{GeoMST} which combined \wspd and \bcp with the Kruskal's algorithm. The
algorithm was improved further by computing some \bcp lazily or avoiding them
altogether.
Recently, Wang et.al~\cite{wang2021fast} developed a parallel shared-memory
variant based on this approach. To our knowledge, this is currently the fastest
sequential and multithreaded parallel open-source implementation. The algorithm
proposed in~\cite{wang2021fast} algorithm was also shown to work with certain
non-Euclidean distance metrics, such as the mutual-reachability distance for
computing \hdbscan~\cite{campello2015hdbscan}.

March et al \cite{march2010dualtreeMST} proposed an \emst algorithm based on
the dual-tree framework. Unlike the single tree algorithm of Bentley and
Friedman, where the nearest neighbor queries are performed separately for every
point, the dual-tree algorithm performs such a query for a subtree in the
spatial search tree. The
algorithm used the component-wide upper and lower distance bounds during the
tree traversal to avoid unnecessary distance computations. Under certain
assumptions on the distribution of points, dual-tree has the best worst-case
asymptotic complexity. In \cite{mcinnes2017}, researchers used the algorithm
for non-Euclidean mutual-reachability distance of the \hdbscan algorithm.

\textbf{Kokkos.}
Kokkos~\cite{kokkos2014,kokkos2022} is a performance-portable programming
model. It provides abstractions for expressing several parallel execution
patterns such as \texttt{parallel\_\{for,reduce,scan\}}. These patterns take
function objects (\emph{e.g.}, C++ lambdas) as arguments to execute for a given
kernel index. While fairly restricted, this programming models allows maximum
flexibility for mapping the patterns to an execution model. To this end, Kokkos
provides an \emph{execution space} abstraction that represents an execution
resource, and a \emph{memory space} that represents an abstract memory
resource. A user is required to make sure that an execution space has access to
the memory space that the data is in. For example, if the execution space is
\texttt{Kokkos::Cuda} and the data is on the host (\texttt{Kokkos::HostSpace}
memory space), an explicit data transfer is required to put the data on the
device (\texttt{Kokkos::CudaSpace} memory space).
% For example, when running an algorithm on Cuda, a user chooses
% \texttt{Kokkos::Cuda} as an ExecutionSpace, and \texttt{Kokkos::CudaSpace} as
% MemorySpace during the compilation.

Kokkos also provides an abstraction for a multi-dimensional array data
structure called \texttt{View}. It is a polymorphic structure, whose layout
depends on the memory the data resides in (host or device). For example, a
one-dimensional view on a GPU would automatically result in a coalesced data
access pattern.

Together, these abstractions are implemented in a C++
library\footnote{\url{https://github.com/kokkos/Kokkos}}. Kokkos supports
multiple backends, allowing the code written in Kokkos to run on a variety of
hardware. Pertinent to this work, Kokkos supports Nvidia GPUs through the
\texttt{Cuda} backend, AMD GPUs through the \texttt{HIP} backend, serial host
through the \texttt{Serial} backend, and parallel host through the
\texttt{OpenMP} backend.

\textbf{ArborX.}
\arborx~\cite{arborx2020} is a performance-portable geometric search library
based on Kokkos. At its core, \arborx implements a highly efficient parallel
data structure, bounding volume hierarchy (BVH), to allow fast computation of
the two types of the search queries: spatial (\emph{e.g.}, searching for all objects
within a certain distance of an object of interest) and nearest (\emph{e.g.},
searching for a certain number of the closest objects regardless of their
distance from an object of interest).

\arborx implements a linear BVH structure following the
works~\cite{karras2012, apetrei2014}, which has been shown to perform well for
low-dimensional data on GPUs. The user data is linearized using a space-filling
curve (Z-curve) to improve the locality of the geometric objects during the
construction. It is then followed by a fully parallel bottom-up construction
algorithm to produce a binary tree structure (hierarchy). Given $n$ data
points, the resulting tree would have $n-1$ internal nodes and $n$ leaf nodes,
for a total of $2n-1$ nodes. This very fast construction algorithm produces a
tree of sufficient quality in most situations.

During the search (also called a traversal), each thread is assigned a single
query, and all the traversals are performed independently in parallel in a
top-down manner. To reduce the data and thread divergence, the queries are
pre-sorted with the goal to assign neighboring threads the queries that are
geometrically close.

% \begin{figure}[htb]
%   % \centering
%   \includegraphics[width=\columnwidth]{figs/candidate-edges}
%   \caption{ \label{fig:candidate-edges} Candidate edges}
% \end{figure}

% \begin{figure}
%   \centering
%   \includegraphics[width=\columnwidth]{figs/merge-comp.pdf}
%   \caption{ \label{fig:merge-comp} Merging components in parallel to form a larger component} 
%   \end{figure}
\section{Algorithm}\label{s:algorithm}

Our algorithm follows the general approach of combining a classical \mst
algorithm with an efficient data structure for finding nearest neighbors for
the components. We use the \boruvka's algorithm as it exposes the most
parallelism out of all classical algorithms (see~\Cref{s:background}).

In this work, we use a single tree algorithm for two reasons.
First, a parallel implementation of the dual-tree algorithms on GPU
accelerators is an open research problem, and a high-performance implementation
is a significant challenge.
Second, a single-tree approach is much easier to implement by reusing efficient
parallel geometric search algorithms, allowing to tap into existing efficient GPU
implementations.
As we will demonstrate in~\Cref{s:results}, a single tree implementation works
well in practice.

\newif\ifcpp % use C++ code instead of flowchart
% \cppfalse
\cpptrue
\ifcpp
\begin{figure}[t]
\begin{lstlisting}[language=C++]
// ExecutionSpace is the Kokkos execution space
// (where a kernel is executed). MemorySpace is
// the Kokkos memory space (where the data resides).
ExecutionSpace exec_space;
Kokkos::View<int*, MemorySpace> labels("labels", n);

// Initialize labels by placing each vertex into a
// separate component
Kokkos::parallel_for(
  Kokkos::RangePolicy<ExecutionSpace>(exec_space, 0, n),
  KOKKOS_LAMBDA(int i) { labels(i) = i; });

// Construct BVH
ArborX::BVH<MemorySpace> bvh(exec_space, data);

// Perform Boruvka iterations
int num_components = n;
do {
  // Propagate leaf node labels to the internal nodes
  // [parallel_for]
  reduceLabels(exec_space, bvh, labels);

  // Compute upper bounds on the length of the shortest
  // outgoing edge for each component [parallel_for]
  Kokkos::View<float*, MemorySpace> upper_bounds =
    computeUpperBounds(exec_space, bvh, labels);

  // Find the shortest outgoing edge for each component
  // [parallel_for]
  Kokkos::View<Kokkos::Pair<int,int>*, MemorySpace>
    component_out_edges =
        findComponentsOutgoingEdges(exec_space, bvh,
                                    labels, upper_bounds);

  // Merge components using the found edges through
  // updating the labels [parallel_for]
  num_components =
    mergeComponents(space, component_out_edges, labels);
} while (num_components > 1);
\end{lstlisting}
  \caption{The single-tree \emst algorithm C++ implementation using ArborX and
  Kokkos.\label{a:emst_arborx}}
\end{figure}
\else

\begin{figure}
\tikzstyle{decision} = [diamond, draw, fill=blue!5,
    text width=5em, text badly centered, inner sep=0pt]
\tikzstyle{block} = [rectangle, draw, fill=blue!5,
    text width=15em, text centered, rounded corners, minimum height=2em]
\tikzstyle{line} = [draw, -latex']
    \begin{center}
  \begin{minipage}{0.65\columnwidth}
\begin{tikzpicture}
    % Place nodes
    \node [block] (bvh) {Construct BVH from data};
    \node [block, below of=bvh, node distance=1.1cm] (init_labels) {Initialize labels\\\textbf{[parallel\_for]}};
    \node [block, below of=init_labels, node distance=1.2cm] (reduce_labels) {Propagate labels to internal nodes\\\textbf{[parallel\_for]}};
    \node [block, below of=reduce_labels, node distance=1.6cm] (upper_bounds) {Compute upper bounds on the length of the shortest outgoing edge for each component\\\textbf{[parallel\_for]}};
    \node [block, below of=upper_bounds, node distance=1.8cm] (outgoing) {Find the shortest outgoing edge for each component\\\textbf{[parallel\_for]}};
    \node [block, below of=outgoing, node distance=1.6cm] (merge) {Merge components using the found edges (update the labels)\\\textbf{[parallel\_for]}};
    \node [decision, below of=merge, node distance=2.1cm] (decide) {Only one \\component?};
    \node [right of=decide, node distance=2cm] {no};
    \node [block, below of=decide, node distance=1.8cm] (stop) {Stop};
    % Draw edges
    \path [line] (bvh) -- (init_labels);
    \path [line] (init_labels) -- (reduce_labels);
    \path [line] (reduce_labels) -- (upper_bounds);
    \path [line] (upper_bounds) -- (outgoing);
    \path [line] (outgoing) -- (merge);
    \path [line] (merge) -- (decide);
    % \path [line] (decide) -- node [right of=decide, node distance=2cm] {no} (reduce_labels);
    \path [line] (decide) edge [in=0, out=0] (reduce_labels);
    \path [line] (decide) -- node[auto] {yes}(stop);
\end{tikzpicture}
  \end{minipage}
    \end{center}
  \caption{The single-tree \emst algorithm.}\label{a:emst_arborx}
\end{figure}
\fi

The \boruvka's algorithm is iterative in nature (see~\Cref{s:background} for an
overview). \Cref{a:emst_arborx} provides a high level overview of our
implementation, with a detailed description provided later in this Section.
Each iteration consists of two phases. In the first phase, we find the shortest
outgoing edge for each component\ifcpp
\\(\texttt{findComponentsOutgoingEdges})\fi. Using these edges, the components
are merged in the second phase\ifcpp~(\texttt{mergeComponents})\fi. We will now
describe the algorithms for both phases.

\subsection*{Finding the shortest outgoing edge}

We will denote by $C_i^k$ the $i$th component on the $k$th \boruvka iteration.
At the start of the \boruvka's algorithm, each component is initialized with an
individual vertex, $C^0_i = \{v_i\}$. As the algorithm proceeds, the
components are merged together using the found edges.

Let $\mathcal{C}^k = \{C^k_i\}_{i=1}^{s_k}$ be the set of components on
iteration $k$, $C^k_i \cap C^k_j = \emptyset$ for $i \neq j$. The goal of this
phase of the algorithm is to find edges $e^k_i, i = 1, \dots, s_k$ such that
$e^k_i = \argmin\{\|(u^k_i, v^k_i)\| \; |\;  u^k_i \in C^k_i \,\mbox{and}\,
v^k_i \in C^k_j, j \neq i\}$. The component $C^k_j$ is the closest component to
$C^k_i$, and we denote this relationship by $C^k_i \rightarrow C^k_j$.

This problem can be seen as the nearest neighbor problem with an additional
constraint that the nearest neighbor of a point must belong to a component
different from the one the point belongs to, followed by choosing the shortest
edge for all points in a component. Thus, we can follow a general approach to
solving nearest neighbor problem on GPUs. Using \arborx, this is done by
assigning each point to a single thread and executing the neighbor searches in
bulk (i.e., with all threads launching at the same time). Each thread executes
a stack-less top-down traversal.

One of the challenges in designing an efficient algorithm lies in that the
components grow in size with each \boruvka iteration. Examining all nearest
points regardless of their component membership becomes progressively more
expensive. Without trimming the number of the distance computations, this leads
to $\bigo{n^2}$ cost on the later \boruvka iterations.

Thus, we propose two optimization procedures to maintain a moderate cost of
each tree traversal regardless of the component size.

\textbf{Optimization 1: subtree skipping.}
We focus on reducing the number of the tree nodes encountered
during the traversal by each thread. Specifically, individual thread skips the
subtrees where each leaf node belongs to the same component as the point
assigned to the thread. A similar approach was proposed in~\cite{mcinnes2017} in
the context of dual-trees. While the benefit of this approach is limited on the
earlier iterations of the algorithm, when the components are small, it is
critical on the later iterations. In our experience, the cost of \boruvka's
iterations tends to progressively decrease, with later iterations typically
taking a small fraction of the earlier ones.

Our implementation uses a flat array of size $n$, called \emph{labels}, to
indicate a membership of a point in a component\footnote{The content of the
array changes on each \boruvka iteration, as the components are merged
together.}. As each point in the dataset is also a leaf in the constructed
tree, we can associate each leaf node with a label of its component.

\begin{figure}
  \centering
  \resizebox{0.45\textwidth}{!}{%
    % Tableau 10 light
\definecolor{Color1}{HTML}{C0C7B8}
\definecolor{Color2}{HTML}{D48C8B}
\definecolor{Color3}{HTML}{E4C6B3}
\definecolor{Color0}{HTML}{ECECEC}
\definecolor{ColorGray}{HTML}{666666}

\begin{tikzpicture}[
    box/.style={rectangle, draw=none, fill=none, text=Color0, text centered, minimum height=0.3cm},
    v_box/.style={rectangle, draw=black, fill=Color0, text=ColorGray, align=center},
    internal_parent/.style={circle, draw=black, text=black},
    internal_node/.style={circle, draw=none, fill=none, text=black},
    leaf_node/.style={circle, draw=black, text=black},
    text centered,
    font=\bf,
    anchor=north west,
    level distance=0.5cm,
    growth parent anchor=south,
    line width=1pt,
    >=stealth,
    tips=proper
  ]
  %%%%%%%%%%%%%%%%%%%%%%%%%%%%%%%%%%%%%%%%
  % Internal boxes
  {[on background layer]
    \node(box_0)[box, minimum width=11.0cm] at (0.0, 0){};
    \node(box_3)[box, minimum width= 5.0cm] at (0.0,-2){};
    \node(box_4)[box, minimum width= 5.0cm] at (6.0,-1){};
    \node(box_1)[box, minimum width= 2.0cm] at (0.0,-3){};
    \node(box_2)[box, minimum width= 2.0cm] at (3.0,-3){};
    \node(box_5)[box, minimum width= 3.5cm] at (7.5,-2){};
    \node(box_6)[box, minimum width= 2.0cm] at (7.5,-3){};
  }
  % Internal nodes
  \node(internal_node_0)[internal_node, left =-0.3cm of box_0] {};
  \node(internal_node_3)[internal_node, right=-0.3cm of box_3] {};
  \node(internal_node_4)[internal_node, left =-0.3cm of box_4] {};
  \node(internal_node_1)[internal_node, right=-0.3cm of box_1] {};
  \node(internal_node_2)[internal_node, left =-0.3cm of box_2] {};
  \node(internal_node_5)[internal_node, left =-0.3cm of box_5] {};
  \node(internal_node_6)[internal_node, right=-0.3cm of box_6] {};
  % Leaf nodes
  \node(leaf_node_0)[leaf_node, fill=Color1] at ( 0.0,-4) {0};
  \node(leaf_node_1)[leaf_node, fill=Color1] at ( 1.5,-4) {1};
  \node(leaf_node_2)[leaf_node, fill=Color1] at ( 3.0,-4) {2};
  \node(leaf_node_3)[leaf_node, fill=Color1] at ( 4.5, -4) {3};
  \node(leaf_node_4)[leaf_node, fill=Color2] at ( 6.0, -4) {4};
  \node(leaf_node_5)[leaf_node, fill=Color3] at ( 7.5, -4) {5};
  \node(leaf_node_6)[leaf_node, fill=Color3] at ( 9.0, -4) {6};
  \node(leaf_node_7)[leaf_node, fill=Color2] at (10.5, -4) {7};
  % Internal parent nodes
  \node(internal_parent_0)[internal_parent, left=0.4cm of internal_node_1, fill=Color1] {1};
    \draw  (internal_parent_0) edge (leaf_node_0);
    \draw  (internal_parent_0) edge (leaf_node_1);
  \node(internal_parent_2)[internal_parent, right=0.4cm of internal_node_2, fill=Color1] {2};
    \draw  (internal_parent_2) edge (leaf_node_2);
    \draw  (internal_parent_2) edge (leaf_node_3);
  \node(internal_parent_5)[internal_parent, left=0.4cm of internal_node_6, fill=Color3] {6};
    \draw  (internal_parent_5) edge (leaf_node_5);
    \draw  (internal_parent_5) edge (leaf_node_6);
  \node(internal_parent_1)[internal_parent, left=1.9cm of internal_node_3, fill=Color1] {3};
    \draw  (internal_parent_1) edge (internal_parent_0);
    \draw  (internal_parent_1) edge (internal_parent_2);
  \node(internal_parent_6)[internal_parent, right=1.9cm of internal_node_5, fill=Color0] {5};
    \draw  (internal_parent_6) edge (internal_parent_5);
    \draw  (internal_parent_6) edge (leaf_node_7);
  \node(internal_parent_4)[internal_parent, right=0.2cm of internal_node_4, fill=Color0] {4};
    \draw  (internal_parent_4) edge (internal_parent_6);
    \draw  (internal_parent_4) edge (leaf_node_4);
  \node(internal_parent_3)[internal_parent, right=4.9cm of internal_node_0, fill=Color0] {0};
    \draw  (internal_parent_3) edge (internal_parent_1);
    \draw  (internal_parent_3) edge (internal_parent_4);
\end{tikzpicture}
  }
  \caption{Propagation of leaf node labels to the internal nodes. Gray denotes
  invalid labels.\label{f:tree_labels_reduction}}
\end{figure}
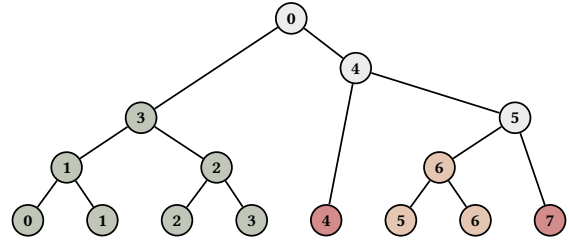

Before running the nearest neighbor algorithm, we propagate the labels
from the leaf nodes to the internal nodes\ifcpp~(\texttt{reduceLabels}
in~\Cref{a:emst_arborx})\fi. For a binary tree-based index, such as our case, this
is done in a single bottom-up traversal algorithm\ifcpp~(\texttt{parallel\_for})\fi. Each thread is
assigned a leaf node, and traverses up the tree. The first thread accessing an internal
node stores its label and terminates, while the second thread combines both
labels, updates the internal node's label and continues upwards. If the labels of
the children of an internal node are the same, the same label is assigned to
that internal node. Otherwise, the internal node is assigned an invalid label
to indicate that the corresponding subtree has leaf nodes from multiple
components. \Cref{f:tree_labels_reduction} shows an example of internal node
labels based on the labels of the leaf nodes. We see that the leaf nodes belong
to three different components, and that there are two subtrees (green and
orange) containing the leaf nodes belonging to a single component. A thread
performing the nearest neighbor search for a point with a green label will skip
the subtree with the root at internal node 3.

\textbf{Optimization 2: upper bounds for the outgoing edges.}
Further improvements may be achieved by using the fact that we are looking for
the closest neighbor for \emph{all} points in a component. The distance to an
encountered neighbor of one point automatically provides an upper bound on the
shortest outgoing edge for the full component. In the extreme case, this upper
bound can be updated each time a thread encounters a leaf node.

However, we take a more moderate approach. We observe that if two points
belong to different components, we can use the distance between them as an
upper bound for the shortest outgoing edge for both of the components. For this
bound to be useful, it is also desirable that these two points are close to each
other. In general, it is not a trivial task to find such pairs. However, one of
the steps in constructing a linear BVH is sorting data along a space-filling
curve (typically, Z-curve using Morton indices). We then use any neighboring
pair of points on the curve with different labels to initialize the upper
bounds for the components\ifcpp~(\texttt{computeUpperBounds}
in~\Cref{a:emst_arborx})\fi. This works well in practice as a pair of points with
close Morton indices are likely to be close geometrically.

\begin{algorithm}[t]
  \caption{Optimized single-thread nearest neighbor traversal algorithm for a given thread with index $i$.\label{a:traversal}}
\begin{algorithmic}[1]
\small
  \State $point \gets data(i)$         \Comment{data point assigned to a thread $i$}
  \State $component \gets labels(i)$   \Comment{component that the point belongs to}
  \State $radius \gets upper\_bounds(component)$ \Comment{cutoff radius}
  \State $shortest\_distance \gets \infty$           \Comment{the best found distance}
  \State $closest\_neighbor \gets \emptyset$           \Comment{the best found neighbor}
  \State Initialize stack with the root node
  \While{stack is not empty}
    \State Pop the stack and assign it to $node$
    \If{$distance(point, node) > radius$}
      \State \textbf{continue}
    \EndIf

    \ForAll{children $child$ of $node$}
      \State $d \gets distance(point, child)$
      \If{$labels(child) \neq component$ and $d \le radius$}
        \If{$child$ is a leaf node}
          \If{$d < shortest\_distance$}
            \State $shortest\_distance \gets d$
            \State $closest\_neighbor \gets child$
            \State $radius \gets d$
          \EndIf
        \Else
          \State{Insert $child$ into the stack}
        \EndIf
      \EndIf
    \EndFor
  \EndWhile
  \State Update the component's shortest outgoing edge if necessary
\end{algorithmic}
\end{algorithm}

\textbf{Traversal algorithm.}
\Cref{a:traversal} shows the pseudo-code of the nearest neighbor algorithm
executed by an individual thread. On line 3, the cutoff radius is set to the
upper bound distance for the component. If the distance from a data point to a
bounding volume of a tree node less than the current value of the cutoff
radius, the children of the node are examined (line 11). We check that there is
at least one node belonging to a different component in the subtree with
$child$ as root, and that the bounding volume of the $child$ node is within the
cutoff distance (line 13). If a child node is a leaf node closer than the
closest neighbor found so far (line 15), we update the cutoff radius value and
the closest neighbor values. Otherwise, if the child is an internal node, it is
inserted into the stack for the later examination (line 20). Finally, once the
closest neighbor this point is found, we compare and update the component's
shortest outgoing edge if necessary (line 21).

\textbf{Non-Euclidean metrics.}
We also note that while we described the procedure for Euclidean distance, it
will also work for certain other metrics. In particular, the mutual
reachability metric used in \hdbscan can be integrated with a regular nearest
neighbor traversal. The only change to the algorithm is that the cutoff radius
during the traversal is set to the mutual reachability distance instead of the
regular Euclidean distance. This is made possible by the fact that the mutual
reachability distance is always greater or equal to the Euclidean one, and thus
nodes truncated by the mutual reachability distance will also be truncated by
the Euclidean distance.

\subsection*{Merging components together}

In the second phase, we use the edges found in the first phase to merge
components together. As mentioned in~\Cref{s:background}, a single iteration of
the \boruvka's algorithm results in chains of components. The merge procedure
is straightforward and is embarrassingly parallel. For every point, we follow
the chain until reaching the terminal pair of components with their shortest
outgoing edges pointing to each other, and update the value of the labels array
to be the component with the smallest index of that pair.

\section{Experimental results}\label{s:results}

\begin{figure*}
   \includegraphics[width=\textwidth]{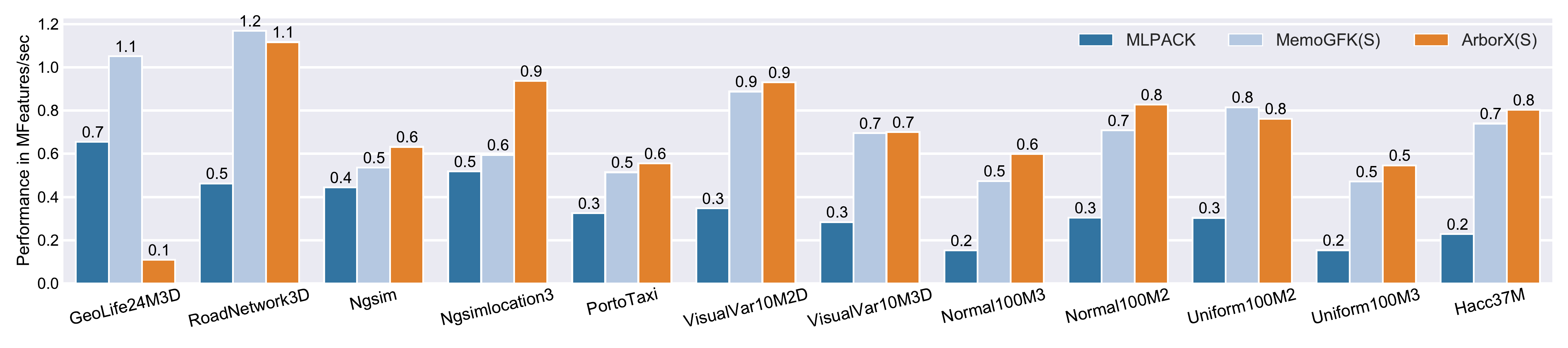}
   \caption{\label{fig:serial-perf-perlmutter} Performance comparison of
   the sequential \emst implementations on \amdcpu.}
\end{figure*}

\begin{figure*}
   \includegraphics[width=\textwidth]{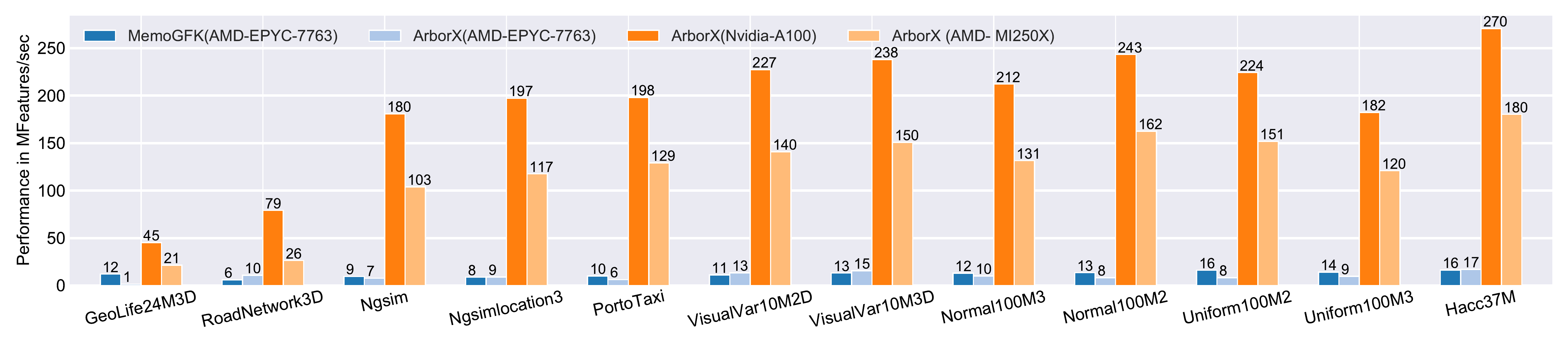}
   \caption{\label{fig:parallel-perf-perlmutter} Performance comparison of
   the parallel \emst implementations using \amdcpu, \nvidiagpu and \amdgpu (single GCD).}
\end{figure*}

In our implementation, we used ArborX~\cite{arborx2020}, an open-source library
for the tree-based implementations, and Kokkos library~\cite{kokkos2022} for
a device-independent programming model.
\unless\ifblind
The implemented algorithm is available in the main ArborX
repository\footnote{\url{https://github.com/arborx/ArborX}}.
\fi

For our rate metric, we used the number of features processed per second,
$nd/t$, where $n$ is the number of points in the dataset, $d$ is the
dimension, and $t$ is the time taken. We also denote by\MilFeatPerSec{}
the number representing millions of features processed per second. We chose to
include the dimension in our rate metric to allow cross-dimensional comparison
of the datasets.

\noindent\textbf{Testing environment.}
The numerical studies presented in the paper were performed using \amdcpu (64 cores),
\nvidiagpu and a single GCD (Graphics Compute Die) of
\amdgpu\footnote{Currently, HIP (Heterogeneous-computing Interface for
Portability) -- the programming interface provided by AMD -- only allows the use
of each GCD as an independent GPU.}. The chips are based on TSMC's N7+, N7 and
N6 technology, respectively, and can be considered to belong to the same generation.

We used GCC 11.2.0 compiler for \amdcpu, NVCC 11.5 for \nvidiagpu, and ROCm 4.5 for \amdgpu.

\noindent\textbf{Datasets.}
For our experiments, we used a combination of artificial and real-world datasets:
\begin{itemize}
  \item \textbf{Ngsim} (2D) ~\cite{ngsim} consists of $\sim$12M 2D points
    corresponding to car trajectories on three highways.
    We also use one of these highways as a separate dataset \textbf{Ngsimlocation3}.

  \item \textbf{PortoTaxi} (2D) ~\cite{portotaxi} consists of 1,710,000+
    trajectories with $\sim$81M 2D points in total, corresponding to the
    trajectories of several hundred taxis operating in the city of Porto,
    Portugal.

  \item \textbf{RoadNetwork3D} (2D) ~\cite{3droad} consists of $\sim$400K 2D points
    of the road network of the North Jutland province in Denmark.

  \item \textbf{GeoLife24M3D} (3D) ~\cite{zheng2008geolife} consists of $\sim$24M 3D
    points corresponding to a user location data (longitude, latitude,
    altitude), and has a very skewed distribution.

  \item \textbf{Hacc37M} (3D) and \textbf{Hacc497M} (3D) consist of the 3D data taken from a single
    rank of a cosmology simulation performed with HACC~\cite{hacc}.
    \dataset{Hacc37M} was taken from a $1024^3$ particles simulation, and has $\sim$37M points. \dataset{Hacc497M} was taken from a $3072^3$ particles simulations, and has $\sim$497M points.

  \item \textbf{VisualVar10M2D} (2D) and \textbf{VisualVar10M3D} (3D) were produced by the generator
    of~\cite{gan2017}. Both datasets are of size 10M.

  \item \textbf{Normal100M2} (2D), \textbf{Normal300M2} (2D), \textbf{Normal100M3} (3D), (\textbf{Normal200M3} (3D)
    consist of randomly generated points with zero mean and one standard deviation in all the dimensions.
    The dataset sizes are 100M, 300M, 100M, 300M, respectively.

  \item \textbf{Uniform100M2} (2D) and \textbf{Uniform100M3} (3D) are randomly generated datasets where all the points are distributed \emph{uniformly}  inside a unit square (cube) in 2D (3D), both centered at the origin. Both datasets are of size 100M.
\end{itemize}

\paragraph{Competing Algorithms:}
We compare the performance of our algorithm to \mlpack~\cite{mlpack2018} implementation of the dual-tree algorithm~\cite{march2010dualtreeMST} available at \url{https://github.com/mlpack/mlpack}, and to \wangemst implementation of the~\cite{wang2021fast} available at \url{https://github.com/wangyiqiu/hdbscan}.

\subsection{Sequential performance}
\label{sec:serial-perf}
Our first goal is to compare the sequential performance of the implementations.
\Cref{fig:serial-perf-perlmutter} shows the results comparing \mlpack,
\wangemst and \arborx on a variety of datasets using a single thread on
\amdcpu.

We observe that \mlpack is slower than \wangemst for all the datasets. The
sequential performance of our algorithm is competitive for most datasets, and
is 1.5$\times$ faster than \wangemst for the \dataset{Ngsimlocation3}. The only
outlier is the \dataset{GeoLife24M3D}. Our investigation showed that the properties
of that dataset make it challenging to construct a high quality BVH.
Specifically, the extremely high density of certain regions is under-resolved by
the space-filling curve, resulting in significant bounding volume overlaps
among nodes of certain subtrees. We believe that this issue can be addressed by
increasing the resolution of the Z-curve grid, e.g., by using 128-bit Morton
codes instead of 64-bit ones.

An interesting observation is that the performance of all implementations seem
to be dimension-agnostic, as the rates are similar between 2D and 3D datasets.

% \item MLpack is the slowest in all but one case. But in most cases, its
% performance is within 50\% of the best performing one. Readers should note that
% MLpack is focused on broader functionality and algorithm with theoretical
% guarantees rather than high-performance implementation.

\begin{figure*}
\subfloat[Hacc497M\label{fig:samplingHacc}]{\includegraphics[width=0.32\textwidth]{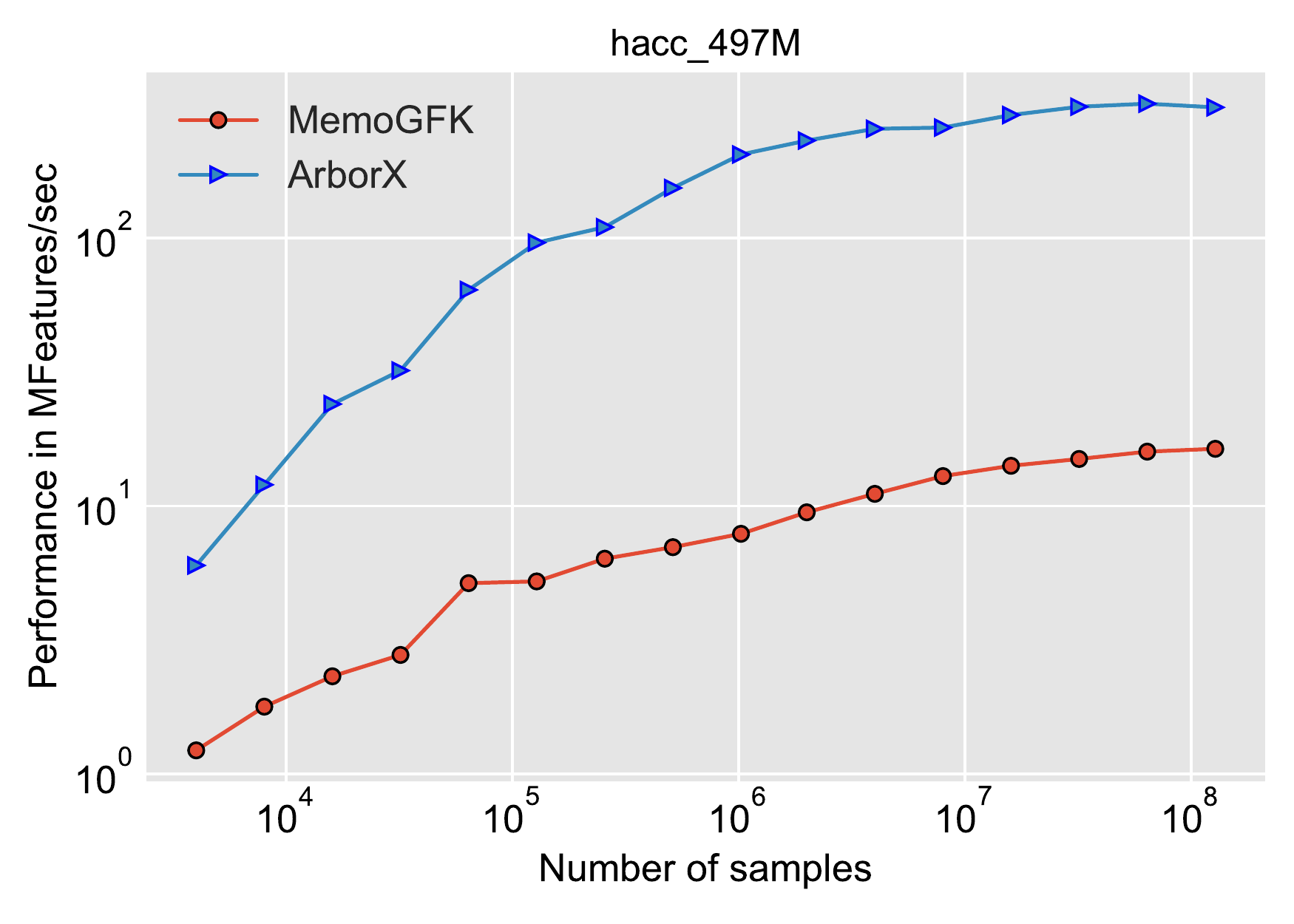}}
\hfill
\subfloat[Normal300M2\label{fig:samplingNormal}]{\includegraphics[width=0.32\textwidth]{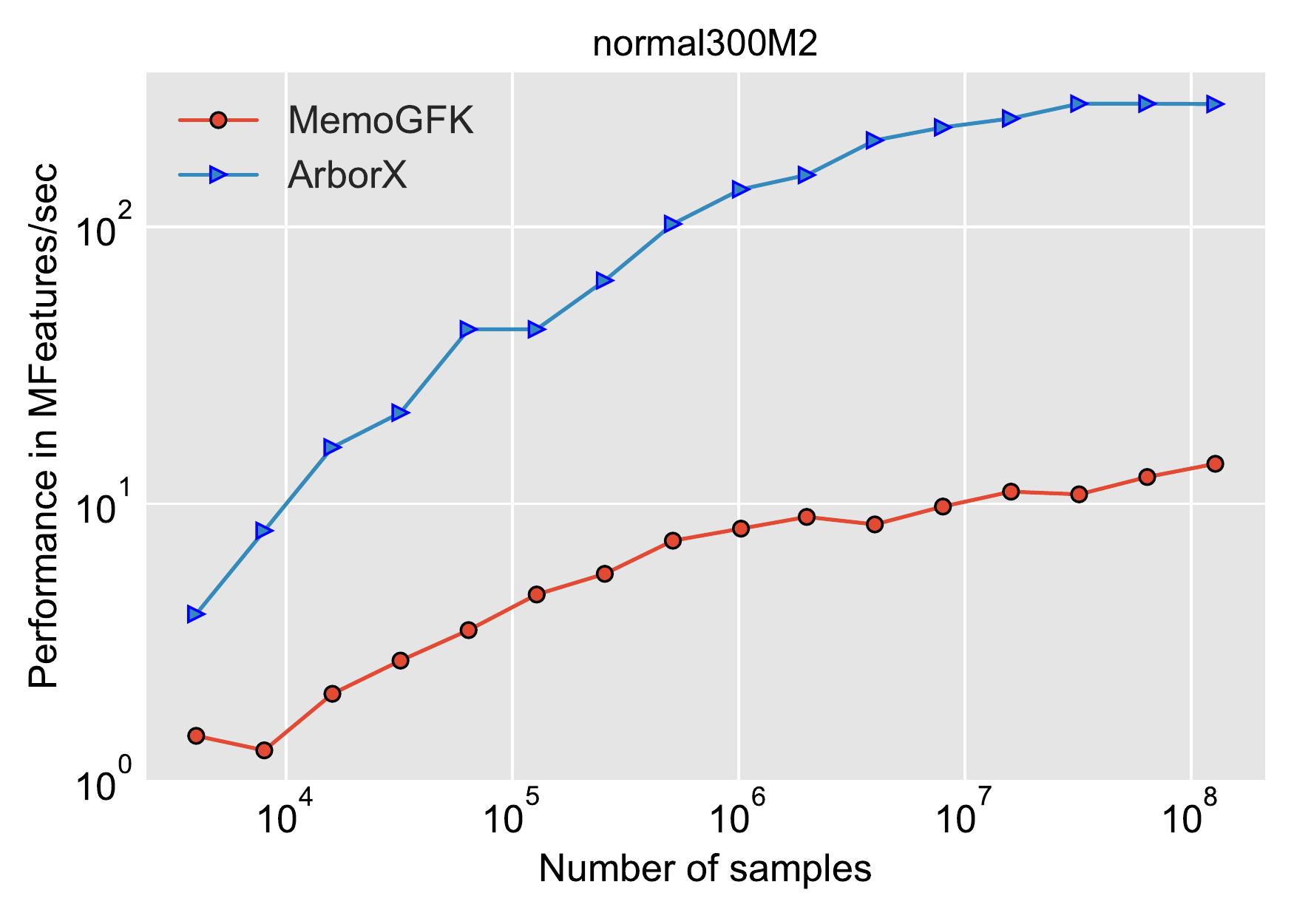}}
\hfill
\subfloat[Uniform300M3\label{fig:samplingUniform}]{\includegraphics[width=0.32\textwidth]{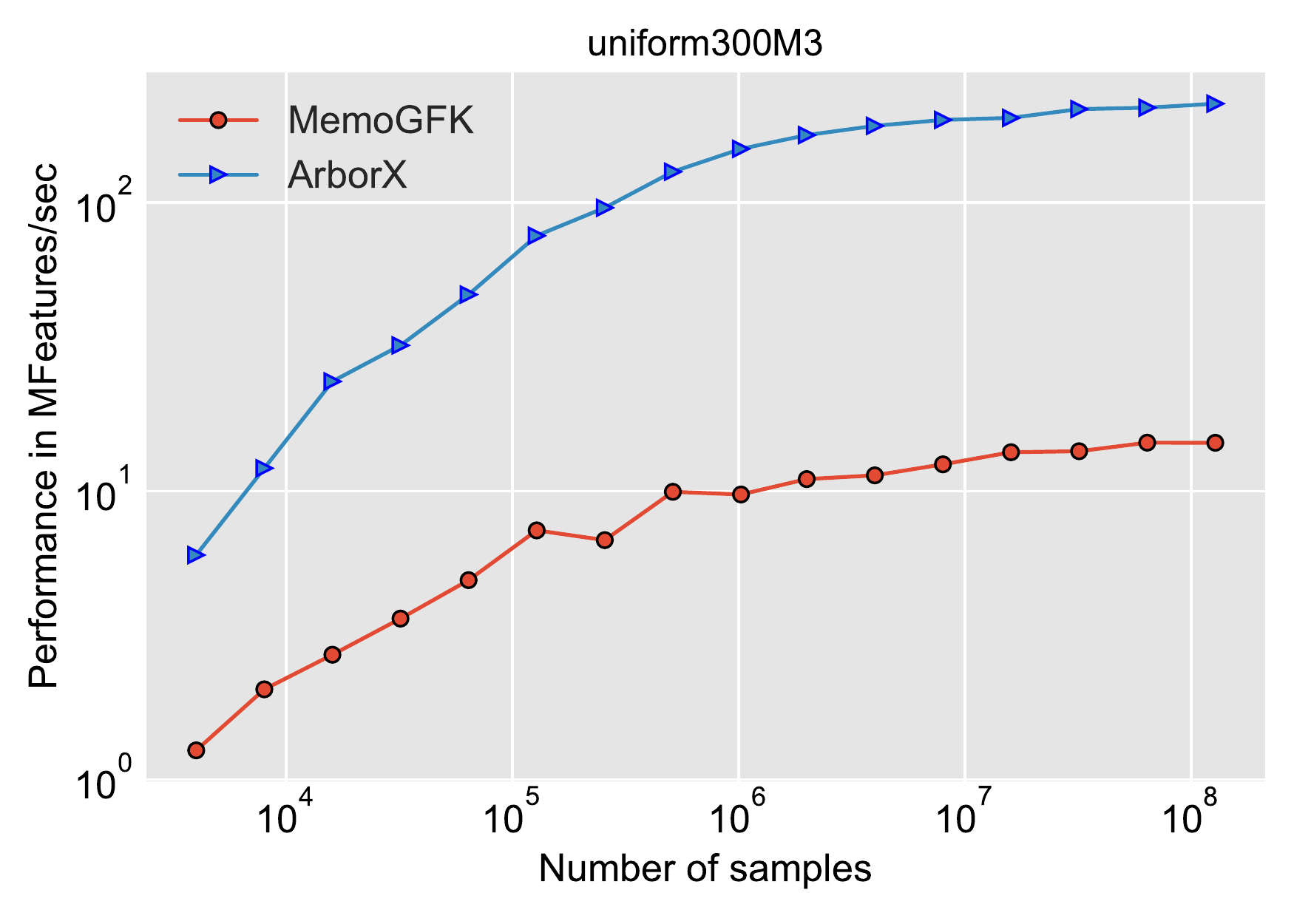}}
\caption{\label{fig:sampling} Effect of the dataset size on the parallel performance using \amdcpu and \nvidiagpu.}
\end{figure*}

\subsection{Parallel performance}
\label{sec:parallel-perf}
We now compare the parallel performance of the best multi-threaded
implementation~\wangemst using \amdcpu with the parallel CPU and GPU
implementations of \arborx run \amdcpu, \nvidiagpu and \amdgpu (single GCD).
The results are presented on~\Cref{fig:parallel-perf-perlmutter}.

We observe that our \arborx implementation achieves 45-270\MilFeatPerSec{}
 on an \nvidiagpu, and is faster by 4-24$\times$ than
\wangemst.
The relative performance between different datasets observed on \amdgpu is
qualitatively similar to observed performance in \nvidiagpu. For both \amdgpu
and \nvidiagpu, we achieve the best performance for \dataset{Hacc37M}, and the
worst performance for \dataset{GeoLife24M3D}. The \arborx on a single GCD of
\amdgpu is faster by 2-12$\times$ than the multithreaded \wangemst on \amdcpu.

The good and bad cases are the similar between \wangemst and \arborx. Both
implementations achieve the best performance on \dataset{Hacc37M}, and the worst
on \dataset{GeoLife24M3D} (see the discussion in the previous Section). We also
observe lower performance of \arborx on the \dataset{RoadNetwork3D}. This is
caused by the smaller size of that dataset, which is not enough to fully
saturate a GPU.

In general, there is little qualitative differences in performance between 2D
and 3D datasets. In other words, performance has little variability with
respect to the dimension of the data, but is more dependent on the distribution
of points. One exception to that are the uniform datasets, where we see up to
20\% reduced performance for the 3D datasets with respect to the 2D dataset.

We find that our \arborx multi-threaded implementation achieves 10-17
\MilFeatPerSec{} on \amdcpu (with an exception of the
\dataset{GeoLife24M3D} dataset), which puts it within factor 0.5-2$\times$ of
the \wangemst. A currently known limitation of the multi-threaded
implementation is the poor scaling of the sort algorithm. The native
multi-threaded \texttt{Kokkos::BinSort} showed very poor performance on some of
the datasets, and was replaced by an \texttt{std::sort}, a serial sort from the
standard C++ library. For larger datasets, the serial nature of this sort
becomes a dominant cost. We look to replace it with a robust multi-threaded
sort implementation in the future.

We also note that we have not used any architecture-specific optimization for
any device, and that we do not attempt to study the impact of architectural
differences in \arborx performance. Nevertheless, we would like to make a
qualifying remark about relative performance on \amdgpu and \nvidiagpu. We
primarily used the \nvidiagpu for algorithm and software development, debugging
and profiling. Doing so may result in \emph{performance bias} for \nvidiagpu
since our algorithmic design process was guided by performance hotspots
observed on the \nvidiagpu.

\subsection{Scaling performance}
\label{sec:scaling}

We now explore the performance of the algorithms with respect to the number of
points in a dataset. As all algorithms are sensitive to the distribution of points in a dataset, we try to maintain a given distribution by randomly sampling a large dataset a specified number of times, producing
a subset with the same data distribution.

We show the results of the sampling experiment for three datasets in \Cref{fig:sampling}.
The performance of each algorithm increases with the number of samples until it reaches saturation.  This empirically demonstrates the asymptotic linear complexity of
the two algorithms. Otherwise, if the complexity was higher than linear, our metric would decrease with increasing number of samples.

We also observe that \arborx seems to start peaking around $10^6$ mark. In contrast, \wangemst achieves its peak performance at much higher number
of points. This is counter intuitive, as typically CPU algorithms reach peak performance at lower problem sizes compared to similar GPU algorithms.

\subsection{Analysis of computational phases}
\label{sec:breakdown}
% \begin{figure}[t]
%    \includegraphics[width=\linewidth]{plots/wangTimeBreakdown-selected.pdf}
%    \caption{\label{fig:wangTimeBreakdown} Breakdown of different phases of \wangemst and their speed-up over sequential on \amdcpu.}
% \end{figure}

% \begin{figure}[t]
%    \includegraphics[width=\linewidth]{plots/ArborxTimeBreakdown-selected.pdf}
%    \caption{\label{fig:arborxTimeBreakdown} Breakdown of different phases of \arborx and their speed-up over sequential on \nvidiagpu.}
% \end{figure}

\begin{figure*}[t!]
\subfloat[\label{fig:wangTimeBreakdown} Breakdown of different phases of \wangemst and their speed-up over sequential on \amdcpu.]{\includegraphics[height=0.17\textheight]{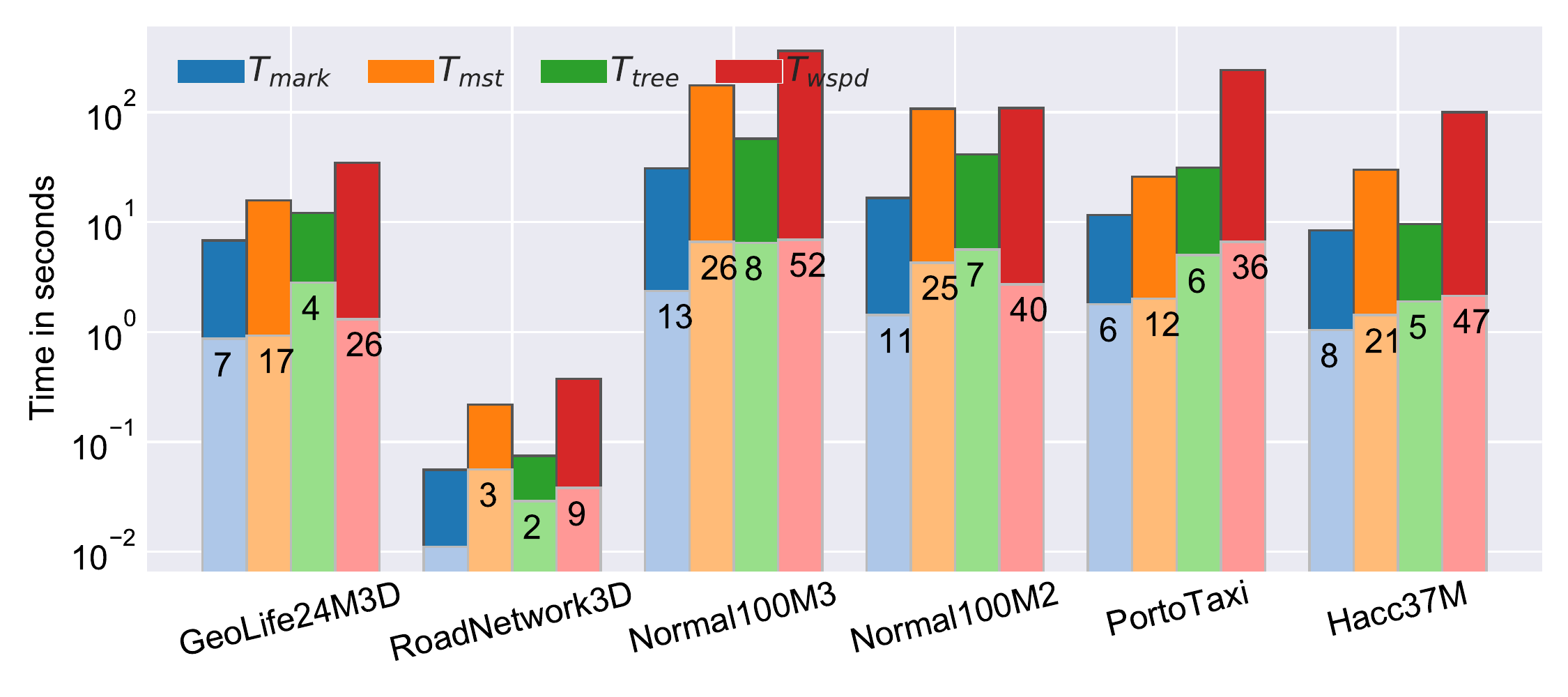}}
\hfill 
\subfloat[\label{fig:arborxTimeBreakdown} Breakdown of different phases of \arborx and their speed-up over sequential on \nvidiagpu.]{\includegraphics[height=0.17\textheight]{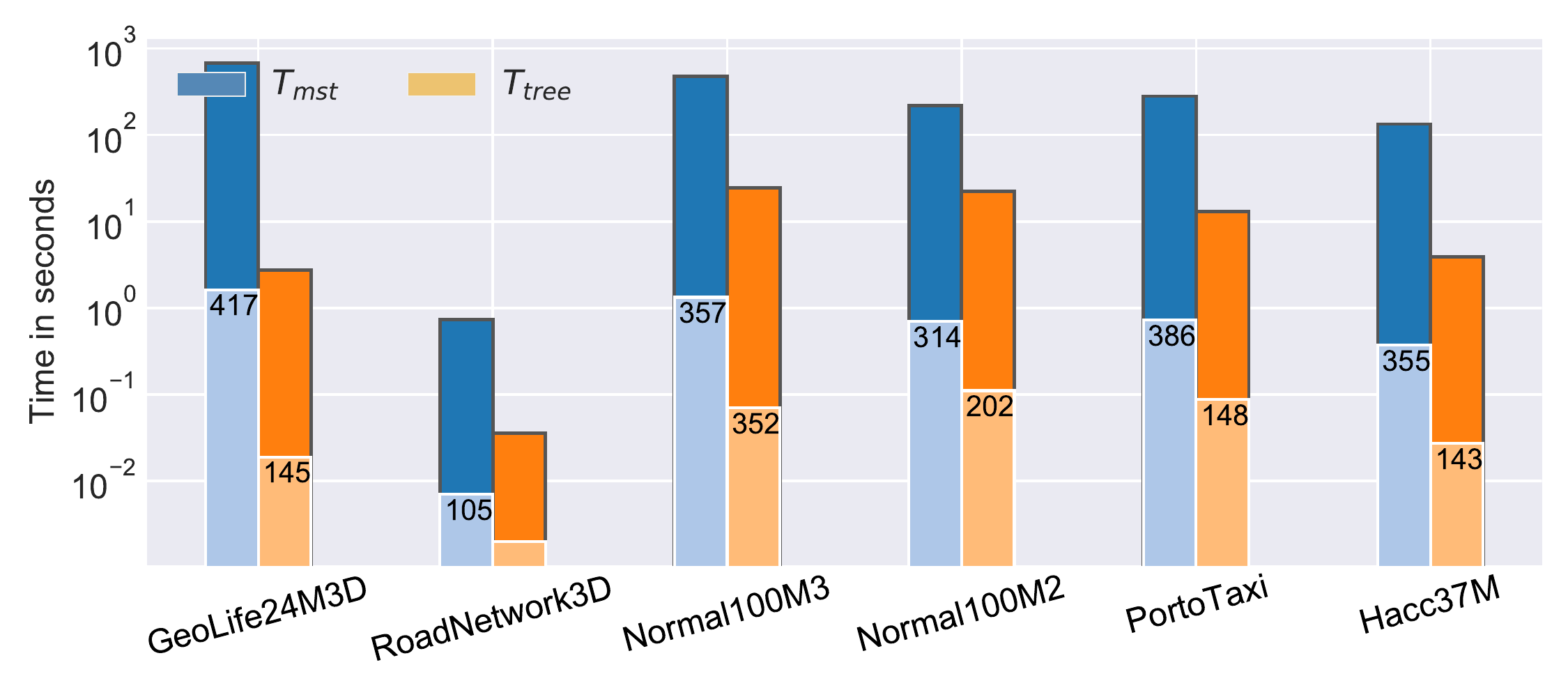}}
\caption{\label{fig:timeBreakDown}Breakdown of different phases of \wangemst and \arborx}
\end{figure*}
We show the relative cost and scaling of the different computation phases for
\wangemst and \arborx algorithms.

\wangemst algorithm consists of four phases:
tree construction ($T_{tree}$),
\wspd calculation ($T_{wspd}$),
Kruskal's \mst algorithm ($T_{mst}$), and
auxiliary routines ($T_{mark}$).
\Cref{fig:wangTimeBreakdown} shows the breakdown for \wangemst. The lower
portion of each bar corresponds to the multi-threaded performance, while the
full bar is the sequential performance. The numbers indicate the ratio between
the two.

We see that in the sequential case, the costliest step is the computation of
\wspd. However, \wspd calculation scales well with the number of cores and
achieves the best case speed-up of 57$\times$ on 64 CPU cores. On the other hand,
tree construction is not a bottleneck in the sequential case, but its poor
scaling makes it the slowest phase of the \emst computation for many datasets.

\arborx algorithm consists of only two phases:
tree construction ($T_{tree}$), and
\boruvka's \mst algorithm ($T_{mst}$).
Except for \dataset{RoadNetwork3D}, which is of small size, both phases scale
well on GPU, and achieve the best speed-up of 360$\times$ and 350$\times$,
respectively.

\subsection{Mutual reachability distance}
\label{sec:mreach}
\hdbscan~\cite{campello2015hdbscan} is a popular unsupervised clustering
algorithm. Similarly to \emst, it seeks to construct an \mst on a complete
graph of a set of points. The main difference is that instead of using
Euclidean distance, it uses the mutual reachability distance (m.r.d.). Given
two points $u$ and $v$, m.r.d. is defined as
$$
  d_{mreach}(u, v) = \max \left\{ d_{core}(u), d_{core}(v), \|u - v\|_2 \right\}.
$$
Here, $d_{core}(u)$ is the \emph{core distance}, defined as the distance to the
$k_{pts}$th nearest neighbor (including the point itself), where $k_{pts}$ is an input
parameter to \hdbscan. When run with $k_{pts} = 1$, $d_{mreach}$ is equivalent to the
regular Euclidean distance.

Computing an \mst in this scenario requires two changes to the regular \emst
calculations. First, core distances have to be determined prior to running an
\mst algorithm. Second, an \emst algorithm must be modified to allow for a
non-Euclidean distance metric. Both \wangemst and \arborx (see~\Cref{s:algorithm}) allow
use of the m.r.d. metric.

\begin{figure*}[!t]
   % \subfloat[\label{fig:coredist-GeoLife24M3D}]{\includegraphics[width=0.25\textwidth]{plots/coredist-GeoLife24M3D.pdf}}
   % \subfloat[\label{fig:coredist-RoadNetwork3D}]{\includegraphics[width=0.25\textwidth]{plots/coredist-RoadNetwork3D.pdf}}
   % \subfloat[\label{fig:coredist-Normal100M2}]{\includegraphics[width=0.25\textwidth]{plots/coredist-Normal100M2.pdf}}
   \subfloat[\label{fig:coredist-Normal100M3}Normal100M3]{\includegraphics[height=.175\textheight]{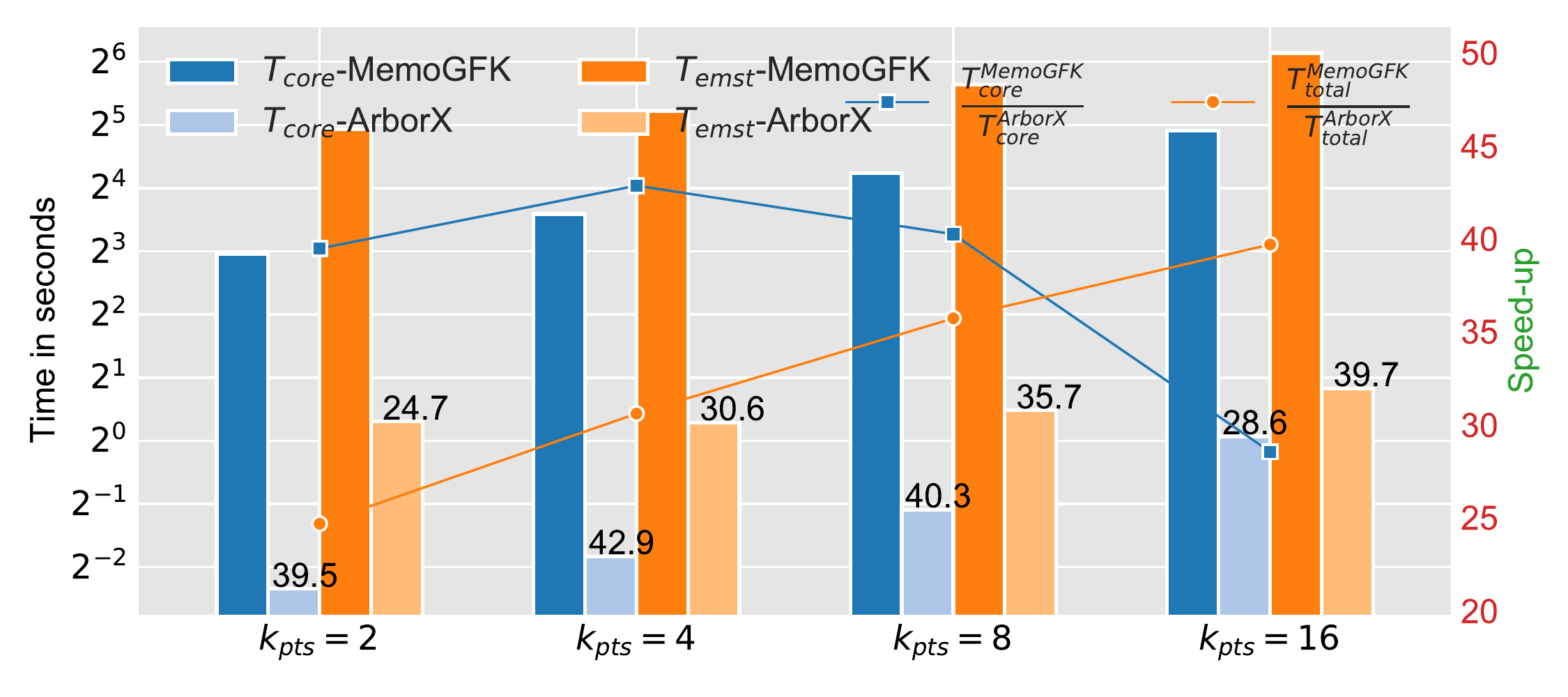}}
  \hfill 
   % \subfloat[\label{fig:coredist-Uniform100M2}]{\includegraphics[width=0.25\textwidth]{plots/coredist-Uniform100M2.pdf}}
   % \subfloat[\label{fig:coredist-Uniform100M3}]{\includegraphics[width=0.25\textwidth]{plots/coredist-Uniform100M3.pdf}}
   \subfloat[\label{fig:coredist-Hacc37M}Hacc37M ]{\includegraphics[height=.175\textheight]{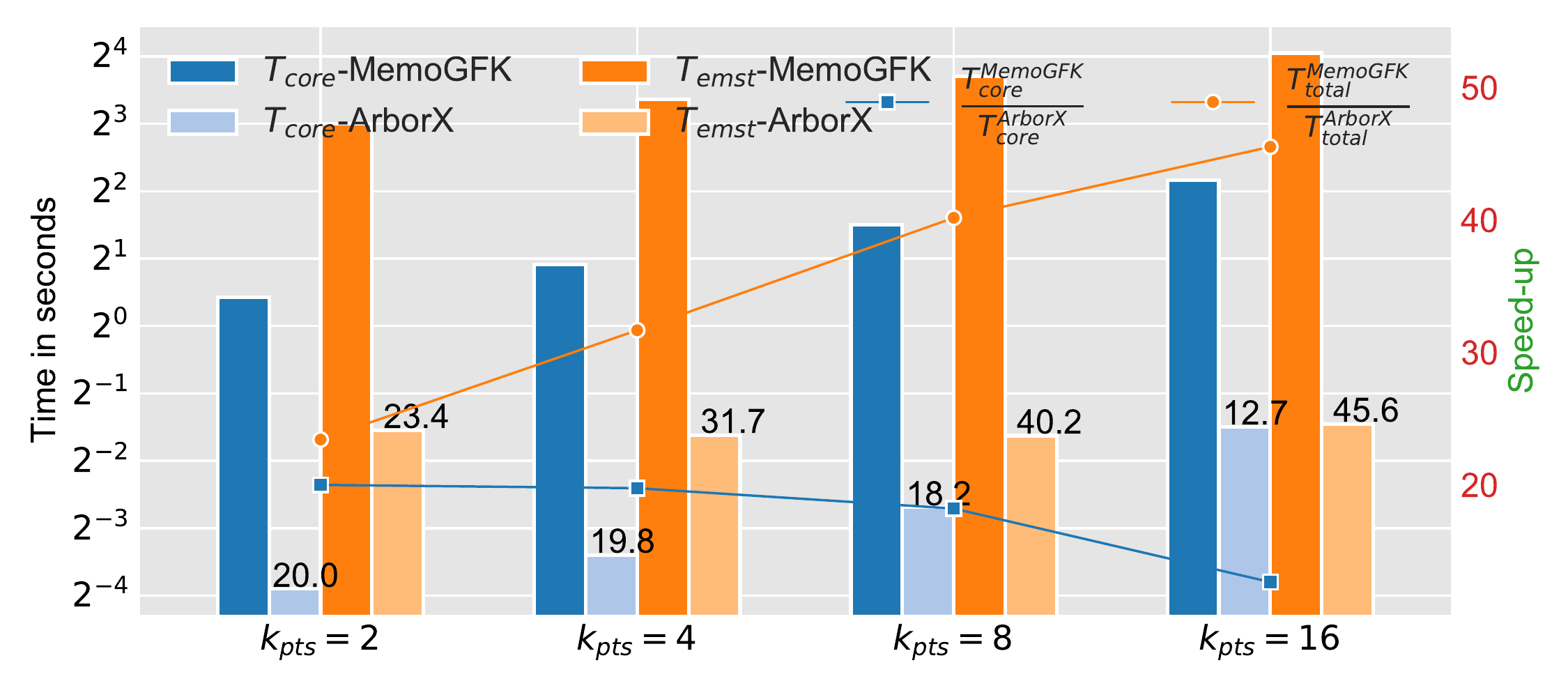}}
\caption{\label{fig:core-distance} Effect of $k_{pts}$ on the \mst performance using mutual reachability distance.}
\end{figure*}

In this Section, we would like to explore the effect of using m.r.d. with
different values of $k_{pts}$ on both the runtime and relative speedup of \wangemst and
\arborx implementations. \Cref{fig:core-distance} shows the effect of varying values
of $k_{pts}$ on the runtime of the implementations for two datasets. $T_{core}$ and $T_{emst}$ denote the
time to compute core-distances and the total time to compute \mst with m.r.d, respectively.

We first observe that increasing values of $k_{pts}$ results in growth of $T_{core}$. This is entirely expected as more neighbors are to be found. However, the kernel cost grows faster in the \arborx implementation on GPU compared to the \wangemst on CPU. For example, for the \dataset{Hacc37M} the speedup of \arborx over \wangemst drops from 20 at $k_{pts} = 2$ to only 12.7 at $k_{pts} = 16$. This is likely caused by the cost of thread divergence when maintaining priority queues for every thread.

The increase in $T_{emst}$ is partially caused by the increase in $T_{core}$. The cost of the \boruvka iterations kernel is less clear. The difference between m.r.d. and Euclidean distance only affects earlier \boruvka iterations, when the distances to the closest neighbors are smaller than their core distances. Thus, many neighbors for a given point will all have the same m.r.d. distance to it, resulting in more expensive neighbor searches. This effect disappears on the later \boruvka iterations when the Euclidean distance dominates. We have also observed that increasing $k_{pts}$ will result in more components getting merged on the earlier \boruvka iterations. In general, the cost of that kernel does not increase much with $k_{pts}$, staying within 30\% of $k_{pts}=2$.

% \begin{figure*}
%    \subfloat[\label{fig:coredist-GeoLife24M3D}]{\includegraphics[width=0.25\textwidth]{plots/coredist-GeoLife24M3D.pdf}}
%    % \subfloat[\label{fig:coredist-RoadNetwork3D}]{\includegraphics[width=0.25\textwidth]{plots/coredist-RoadNetwork3D.pdf}}
%    % \subfloat[\label{fig:coredist-Normal100M2}]{\includegraphics[width=0.25\textwidth]{plots/coredist-Normal100M2.pdf}}
%    \subfloat[\label{fig:coredist-Normal100M3}]{\includegraphics[width=0.25\textwidth]{plots/coredist-Normal100M3.pdf}}
%    %
%    % \subfloat[\label{fig:coredist-Uniform100M2}]{\includegraphics[width=0.25\textwidth]{plots/coredist-Uniform100M2.pdf}}
%    \subfloat[\label{fig:coredist-Uniform100M3}]{\includegraphics[width=0.25\textwidth]{plots/coredist-Uniform100M3.pdf}}
%    \subfloat[\label{fig:coredist-Hacc37M}]{\includegraphics[width=0.25\textwidth]{plots/coredist-Hacc37M.pdf}}
% \caption{\label{fig:core-distance} Performance comparison for different $k_{pts}$ used as core distance for HDBSCAN}
% \end{figure*}

\section{Conclusion}
We presented a single-tree algorithm for the \emst problem designed to exploit
the massively threaded parallelism available on GPUs.
The key strength of our approach is its simplicity through the use of a
single-tree traversal with certain optimizations to prune the neighbor search.
We evaluated the sequential, multithreaded, and GPU versions of our approach
using a variety of datasets on multiple hardware architectures including Nvidia
and AMD GPUs.
We demonstrated that it was performance portable across these platforms and its
excellent performance on GPUs compared to the best multi-threaded
implementation.
We conclude that our approach is efficient for a low-dimensional data.
It remains to be seen if our findings hold for the data of higher dimension,
which we plan to explore in our future work.

% \section*{CRediT author statement}

% \textbf{Andrey Prokopenko}: Conceptualization, Investigation, Software, Writing - original draft.
% \textbf{Piyush Sao}: Investigation, Validation, Software, Writing - review and editing.
% \textbf{Damien Lebrun-Grandi\'e}: Software, Writing - review and editing.

\unless\ifblind
\section*{Acknowledgements}
% ECP disclaimer
This research was supported by the Exascale Computing Project (17-SC-20-SC), a
collaborative effort of the U.S. Department of Energy Office of Science and
the National Nuclear Security Administration.

% OLCF disclaimer
This research used resources of the Oak Ridge Leadership Computing Facility at
the Oak Ridge National Laboratory, which is supported by the Office of Science
of the U.S. Department of Energy under Contract No. DE-AC05-00OR22725.
\fi

\ifjournal
  \bibliographystyle{ACM-Reference-Format}
\else
  \bibliographystyle{apalike}
\fi
\bibliography{boruvka}

\end{document}